\documentclass[]{aa}
\usepackage{amssymb,longtable,graphics}
\usepackage{natbib}

\begin{document}

\title{\textit{Chandra} X-ray spectroscopy of a clear dip in GX 13+1}
 
\author{A. D'A\`i\inst{1}
\and R. Iaria \inst{1}
\and T. Di Salvo \inst{1}
\and A. Riggio \inst{2} 
\and L. Burderi \inst{2} 
\and N.R. Robba \inst{1}
}

\institute{Dipartimento di Fisica e Chimica,  Universit\`a di Palermo, via Archirafi 36 - 90123 Palermo, Italy
\email{antonino.dai@unipa.it} \and
Universit\`a degli Studi di Cagliari, Dipartimento di Fisica, SP Monserrato-Sestu, KM 0.7, 09042 Monserrato, Italy
}

\date{Received 10/06/2013 ; accepted 26/02/2014}

\abstract {The source  GX 13+1 is a persistent,  bright Galactic X-ray
  binary hosting an  accreting neutron star.  It  shows highly ionized
  absorption features, with a blueshift  of $\sim$ 400 km s$^{-1}$ and
  an outflow-mass rate similar to the accretion rate. Many other X-ray
  sources exhibit warm absorption features, and they all show periodic
  dipping behavior at  the same time. Recently,  a dipping periodicity
  has also  been determined for  GX 13+1 using long-term  X-ray folded
  light-curves,  leading to  a  clear identification  of  one of  such
  periodic dips in  an archival $Chandra$ observation.}   {We give the
  first spectral characterization of the periodic dip of GX 13+1 found
  in this  archival \textit{Chandra}  observation performed  in 2010.}
          {We used \textit{Chandra}/HETGS data (1.0--10 keV band) and
            contemporaneous \textit{RXTE}/PCA  data (3.5--25  keV) to
            analyze   the  broad-band   X-ray  spectrum.    We  adopted
            different  spectral  models   to  describe  the  continuum
            emission   and  used   the  XSTAR-derived   warm  absorber
            component  to  constrain  the  highly  ionized  absorption
            features.}   {The  1.0--25   keV  continuum  emission  is
            consistent with  a model  of soft  accretion-disk emission
            and an optically thick, harder Comptonized component.  The
            dip event,  lasting $\sim$  450 s, is  spectrally resolved
            with  an increase  in the  column density  of the  neutral
            absorber, while  we do not find  significant variations in
            the column  density and  ionization parameter of  the warm
            absorber with  respect to  the out-of-dip  spectrum.}  {We
            argue that the very low dipping duty-cycle with respect to
            other sources  of the  same class can  be ascribed  to its
            long orbital period and the  mostly neutral bulge, that is
            relatively small compared with the dimensions of the outer
            disk radius.}

  
\titlerunning{\textit{Chandra} spectroscopy of a clear dip in GX 13+1}
\authorrunning{A. D'A\`i et al.}
\maketitle

\section{Introduction}

Low-mass  X-ray binary  (LMXB)  dipping sources  are characterized  by
periodic  (or  quasi-periodic) dips  in  their  light-curves that  are
evidence for  a fixed structure in  the reference frame of  the binary
system. These dips may also be related to super-orbital periodicities,
which  are  more  difficult  to constrain  when  their  appearance  is
transient \citep{grise13}.  To  date, 13 LMXBs hosting  a neutron star
(NS) and 6 LMXBs  hosting a black hole have shown  clear dips in their
light curves.   In Table~\ref{table_sources}, we show  an updated list
of  these  sources  with  some   basic  data  and  references  to  the
literature.

\begin{table*}
\caption{\label{table_sources} LMXB dipping sources}
\centering
\begin{tabular}{l ccccc}
\hline \hline
\multicolumn{6}{c}{NEUTRON STAR LMXBs}\\
\hline
Source      			&  P$_{\textrm{orb}}$ & M$_2$               & D           & Ref. (dips) & Ref. (WA) \\
            			&  hr                 & M$_{\textrm {sun}}$ & kpc         &      \\
\hline
EXO 0748-676         & 3.82    &    0.1?        & 7.1$\pm$1.2   & 1a & 1b\\
4U 1254-690          & 3.88    &    0.4?        & 10?           & 2a & 2b\\ 
GX 13+1              & 588     &    $>$1.1      & 7$\pm$1       & 3a & 3b \\
4U 1323-62           & 2.94    &    0.3?        & 10?           & 4a & 4b \\
X1624-490            & 20.9    &    2.3?        & 15?           & 5a & 5b \\
X1658-298            & 7.12    &    0.8?        & 15?           & 6a & 6b \\ 
XTE J1710-281        & 3.28    &    0.4?        & 16?           & \multicolumn{2}{c}{7} \\
AX J1745.6–2901      & 8.35    &    0.9?        & 10?           & \multicolumn{2}{c}{8} \\
1A 1744-361          & 0.87?   &    0.1?        & 9?            & 9a & 9b \\
XB 1746-371          & 5.73    &    0.6?        & 9?            & 10a & 10b \\
GRS J1747-312        & 12.4    &    4.5$\pm$3.5 & 6.8$\pm$0.5   & 11 & $\cdots$\\
XB 1916-053          & 0.83    &    0.1?        & 9?            & 12a & 12b \\
Cir X-1              & 400.32  &    10?         & 6?            & 13a & 13b \\
\hline
\multicolumn{6}{c}{BLACK-HOLE LMXBs} \\
\hline
GRO J1655-40\tablefootmark{a}   & 62.92   & 2.34$\pm$0.12 & 3.2$\pm$0.2    & 14a & 14b \\ 
H 1743-322                      & $>$10?  & $\cdots$      & 8.5?           & 15a & 15b \\
GRS 1915+105\tablefootmark{b}   & 739.2   & 0.8$\pm$0.5   & 9.4$\pm$0.2    & 16a & 16b\\
4U 1630-47                      & $\cdots$& $\cdots$      & 10?            & 17a & $\cdots$ \\
MAXI J1659-152                  & 2.414  &  0.20$\pm$0.05&  8.6$\pm$3.7   & 18a & $\cdots$\\
MAXI J1305-704               &  9.74    &   $<$ 1?     &  6?            & \multicolumn{2}{c}{19}\\
\hline
\hline
\end{tabular}

\tablefoot{ The question mark indicates very uncertain values. Columns
  list the most often used source name, orbital period, companion star
  mass, distance, reference to most recent works related to absorption
  dips and  X-ray spectroscopy of warm  absorption features. Companion
  star masses are estimated assuming that the companion belongs to the
  lower   main    sequence.    \tablefoottext{a}{M$_{\textrm{BH}}$   =
    7.0$\pm$0.2} \tablefoottext{b}{M$_{\textrm{BH}}$ = 12.9$\pm$2.4} }

\tablebib{
(1a) \citet{parmar86};
(1b) \citet{vanpeet09};
(2a) \citet{smale02};
(2b) \citet{diaztrigo09}; 
(3a) \citet{iaria14};
(3b) \citet{diaztrigo12};
(4a) \citet{parmar89}; 
(4b) \citet{boirin05}; 
(5a) \citet{smale01};
(5b) \citet{iaria07}; 
(6a) \citet{oosterbroek01};
(6b) \citet{sidoli01};
(7) \citet{younes09}; 
(8) \citet{hyodo09}; 
(9a) \citet{bhattacharyya06};
(9b) \citet{gavriil12};
(10a) \citet{balucinska04};
(10b) \citet{diaztrigo06}; 
(11) \citet{intzand03}; 
(12a) \citet{white82};
(12b) \citet{boirin04};
(13a) \citet{shirey99};
(13b) \citet{dai07a}; 
(14a) \citet{kuulkers98}; 
(14b) \citet{ueda98}; 
(15a) \citet{homan05};
(15b) \citet{miller06};
(16a) \citet{naik01}; 
(16b) \citet{lee02}; 
(17) \citet{kuulkers98}; 
(18)\citet{kuulkers13};
(19)\citet{shidatsu13}.
}
\end{table*} 

Two main  physical models  are widely discussed  in the  literature to
explain the  occurrence of dips: \citet{white82}  proposed a variable,
azimuthal-dependent height  of the  accretion disk's  outer rim  and a
large system-inclination  angle.  According to the  orbital phase, our
line-of-sight  partially or  totally  intercepts the  rim that  causes
local  absorption of  X-rays produced  in the  innermost parts  of the
system.   The  rim  geometry  was  empirically  adjusted  by  matching
synthesized geometries and the regular  patterns observed in the light
curves  \citep[both at  X-rays  and  in the  optical-UV,  e.g. for  4U
  1822-371,][]{mason86}.  Some dippers also show periods without dips,
which  points  to  a  strong  variability  of  the  occulting  regions
\citep{smale99}.  Alternatively,  another explanation was  proposed by
\citet{frank87}: if  matter from  the companion star  is able  to skim
across the thickness of the outer accretion disk, part of the incoming
stream may  impact the disk  at a much closer  radius \citep{lubow89};
when  part of  this  stream  collides with  the  disk,  it is  quickly
dynamically and thermally virialized; but a fraction of it (which is a
tunable parameter of the model)  receives energy from the impact shock
and splits  into a two-zone  medium, forming blobs of  cold, condensed
gas,  surrounded   by  a   lower  density   hotter  plasma   at  large
scale-heights above the disk  \citep{krolik81}.  This scenario is able
to  partially account  for  many  empirical facts  such  as the  dip's
periodic occurrence, the dependence on orbital phase, and the duration
and time scales of the single dips.  Both scenarios involve the common
ingredient of a  high inclination angle. For  low-mass companion stars
with  short orbital  periods  ($<$ 1  day),  the inferred  inclination
angle,  $i$, is  constrained  between  65$^{\circ}$ and  85$^{\circ}$,
while for  higher inclinations  eclipses are  also expected.  In these
eclipsing binaries, direct emission from the NS is blocked by the disk
thickness and  only scattered  emission from an  accretion-disk corona
(ADC) may be observed \citep{iaria13}.

Together with these physical scenarios, many studies have been focused
on  deriving  geometrical  and   physical  constraints  by  spectrally
resolving  the dip  events.  Spectra  from dipping  sources have  been
fitted using  a two-component  spectral decomposition consisting  of a
thermal  black-body  emission  from  the  surface  of  the  NS  and  a
Comptonized emission  (usually fitted with a  cut-off power-law). Seed
photons of the  Comptonized spectrum come from the  accretion disk and
the  Comptonization is  thought to  occur at  large disk  radii in  an
extended corona, whose  radius is $>>$ 10$^{9}$ cm.  Using the ingress
and egress times  of the deep dips (where emission  is totally blocked
at the  dip bottom),  it has  been shown that  the corona  emission is
gradually   covered  (\textit{progressive   covering}  approach)   and
therefore extended,  with a  disk-like geometry, while  the black-body
emission   is   point-like  and   attributed   to   the  NS   emission
\citep{church04}.  The  main assumption in deriving  the estimates for
the ADC radius is  that the dip is caused by the  bulge located at the
outer  accretion disk  \citep[as described  by the  geometry envisaged
  by][]{white82},  whose main  effect is  a progressive  photoelectric
absorption of the primary incident source flux.

Detection in the past decade of resonant absorption features of highly
ionized elements in the X-ray spectrum (see Table~\ref{table_sources})
has provided  new clues for  separating the spectral  formation. Local
absorption features often appear to be blue-shifted, which points to a
disk-wind   or  generally   out-flowing,  photoionized   plasma.   The
ionization state of  the optically thick absorbing  plasma is variable
and the time scales can be as short as a few ks \citep{ueda04}, with a
wind velocity  of thousands of  km/s. In  all cases, the  most clearly
resolved lines are from H-like  and He-like transitions of iron, which
implies that the ionization parameter,  $\xi$, of the warm absorber is
$>$  100  \citep{kallman04}.    \citet{boirin05}  first  advanced  the
hypothesis  that  during  dipping  there  might  be  a  \textit{tight}
relation between  cold and warm  absorption because the  overall X-ray
variability  during dipping  would be  driven by  fast changes  in the
column density and ionization state of the warm absorbing medium along
our line  of sight.   In this scenario,  there is no  more need  for a
partial  covering of  an extended  continuum corona  because the  soft
excess  observed  during  dipping  is naturally  accounted  for  by  a
combination  of strong  increase in  the  column density  of the  warm
medium and  a decrease  of its ionization  parameter. Outside  dips, a
warm absorber has been always  observed, which implies that the medium
has a cylindrical distribution and is not confined to the locus of the
bulge.  In  light of these  findings, the continuum  decomposition has
also been questioned,  because an extended corona was felt  to be less
necessary \citep{diaztrigo12}.

\subsection*{The source GX13+1}

The source  GX 13+1 is a  persistent X-ray binary system  belonging to
the  so-called class  of GX  bright bulge  sources, with  an estimated
distance of 7 $\pm$ 1 kpc.  The compact object is an accreting NS that
has sporadically shown type-I X-ray bursts \citep{matsuba95}, orbiting
an  evolved   mass-donor  giant   star  of   spectral  class   KIII  V
\citep{bandyopadhyay99}.   The  system has  peculiar  characteristics,
being in between the classification of low-mass and high-mass systems;
this is  also testified by the  long orbital period of  $\sim$ 24 days
\citep{corbet10,iaria14},   which  makes   it   the   LMXB  with   the
second-longest orbital  period after GRS  1915+105 (30.8 d  period; it
has a black hole as accreting compact object).

The 3--20 keV \textit{Rossi-XTE}  (\textit{RXTE}) spectrum of GX 13+1
has been investigated by \citet{homan04}, in connection with its radio
emission.    The   spectrum   was   deconvolved   according   to   the
\textit{Eastern}  interpretation,   that  is  the  sum   of  a  softer
multicolored accretion-disk emission and a thermal Comptonized, harder
optically  thick emission  in the  boundary layer.  The low-resolution
\textit{RXTE}/PCA spectrum also needed some local features (broad iron
Gaussian line  and a 9 keV  absorption edge) to obtain  a satisfactory
fit.  More recently, emission of higher  than 20 keV has been observed
with   \textit{INTEGRAL}/ISGRI   data    \citep{paizis06},   and   was
subsequently analyzed according to  a thermal plus bulk Comptonization
model \citep{mainardi10}.

Using narrower  bands such as  the 1--10  keV CCD typical  range, the
general continuum adopted  was found to be well  approximated with the
sum   of   soft  disk   emission   and   black-body  harder   emission
\citep{ueda01,sidoli02,ueda04,diaztrigo12}.   The black-body  emission
approximates an optically thick  Comptonized emission, which we deduce
from the  difficulty in  constraining both the  optical depth  and the
electron temperature with a limited  energy range and the high optical
depths characteristics of the very soft spectra of bright accreting NS
LMXBs.  Analysis of K-$\alpha$ edge depths  has also shown that in the
direction   of  the   source  the   ISM  composition   (or  absorbing,
circumbinary cold  matter) is  significantly overabundant  in elements
heavier  than oxygen  \citep[e.g. silicon  and sulphur]{ueda05},  with
X-ray fine-structure  absorption features (XAFS)  around the Si  and S
K-$\alpha$ edges.

High-resolution spectroscopy with  the \textit{Chandra} HETGS revealed
a radiatively/thermally driven  disk wind with an  outflow velocity of
$\sim$ 400  km s$^{-1}$ and  multiple absorption features  from highly
ionized  elements   \citep{ueda04}.   The  wind  probably   carries  a
significant fraction of the total mass-accretion rate, up to 10$^{18}$
g  s$^{-1}$.  Observations  with \textit{XMM-Newton}  also revealed  a
broad (equivalent width $\gg$ 100 eV) iron emission line, whose origin
is associated to  a disk-reflection component, and the  broad width is
ascribed to Compton  broadening in the warm corona.  A global spectral
account of  the total  variability has also  been proposed,  where the
main drivers for the spectral  variability are neutral cold absorption
and    variability   associated    with    a   reflection    component
\citep{diaztrigo12}.

Periodic dipping in the LMXB GX 13+1 was suspected for a long time, on
the  basis of  an  energy-dependent modulation  observed in  long-term
light     curves     \citep{corbet10,diaztrigo12}.     \citet{iaria14}
systematically  searched  in  archived X-ray  observations  for  clear
signatures of  periodic dips. Applying timing  techniques to long-term
folded X-ray light  curves provided a successful method that  led to a
refined  orbital-period  estimate  (24.5274(2)  d) and  to  the  first
ephemeris for the dip passage times.  The only periodic dip that could
be  assigned on  the  basis  of this  ephemeris  for  a pointed  X-ray
observation was in an  archival \textit{Chandra} observation performed
in 2010.  This  corroborates that inclination is a key  factor to spot
warm absorbing  winds in LMXBs  and that  they are optically  thick to
radiation only close to the plane of the accretion disk. This relation
has recently  also been  pointed out  by \citet{ponti12}  for Galactic
LXMBs hosting black-holes.  We present  in this article the results of
the spectroscopic  analysis of the  $Chandra$ dip event,  showing that
the main driver of  the dipping in this source is  an increase in cold
photoelectric absorption.
%
%
%
\section{Observation and data reduction}

We  used  CIAO  4.5  for  the  \textit{Chandra}  data  extraction  and
analysis, CALDB  4.4.7 for  the calibration  data files,  the software
package   HeaSOFT  version   6.13  for   the  \textit{RXTE}/PCA   data
extraction,  reduction,  and   scientific  analysis,  the  Interactive
Spectral Interpretation  System (ISIS)  1.6.2 \citep{houck02}  for the
spectral fitting, and Xspec  v.12.8.1 for spectral models. Observation
times are given in Coordinated Universal Time (UTC).

\subsection{\textit{Chandra} observation}

The   source   GX~13+1   was   observed  multiple   times   with   the
\textit{Chandra} observatory from 2004 to  2012.  For the present work
we  used  the   observation  with  sequence  number   11814  from  the
\textit{Chandra}  archive.   The  observation  started  at  2010-08-01
00:32:37   UTC  and   lasted  28.12   ks.   At   the  same   time  the
\textit{Rossi-XTE}  satellite  observed  the  source,  which  provided
overlapping, although not complete, monitoring.

The observation performed in (faint) timed exposure mode used the High
Transmission Grating  Spectrometer (HETGS)  to diffract  the spectrum,
and  a (350  rows) subarray  of the  ACIS-S detector  to mitigate  the
effects of photon pile-up, with a CCD frame time of 1.24104 s.
  
The brightness of the source ($\sim$ 6$\times$ 10$^{-9}$ erg cm$^{-2}$
s$^{-1}$) prevented  studying the zeroth-order events  since these are
strongly  affected by  pile-up.  The  location  of the  center of  the
zeroth-order    image    was    therefore   determined    using    the
\textit{tg\_findzo}
script\footnote{\url{http://space.mit.edu/cxc/analysis/findzo/}},   as
recommended  by  the \textit{Chandra}  team  in  case of  zeroth-order
pile-up.   We derived  a source  position  at R.A.  = 18$^{h}$  14$^m$
31$^{s}$.08,  Dec  (J2000)   =  -17$\degr$  09$\arcmin$  26$\arcsec$.1
(J2000,  0.$\arcsec$6 uncertainty),  compatible  with the  coordinates
reported in \citet{iaria14}.  Data  were extracted from regions around
the  medium- and  high-energy  grating  arms (MEG  and  HEG) with  the
\textit{tg\_create\_mask}  tool,  manually  setting  the  zeroth-order
position as derived  by the \textit{tg\_findzo} script  and choosing a
width  for HEG  and  MEG arms  of  25 sky  pixels.  We extracted  only
first-order  positive  (HEG+1,  MEG+1)  and  negative  (HEG-1,  MEG-1)
spectra, as  they provide  the best  signal-to-noise ratio  and higher
orders are mostly affected by pile-up.  We used standard CIAO tools to
derive all the other related spectral products.

Spectra were finally  re-binned to have at least 25  counts per energy
channel  to allow  the use  of $\chi^2$  statistics. We  used the  HEG
spectra  in the  1.0--10.0  keV  range and  the  MEG  spectra in  the
1.0--5.0  keV range;  both spectra  are background-dominated  below 1
keV.

\subsection{\textit{RXTE} observation}

For \textit{RXTE}/PCA  spectra we used source  and background spectra,
matrices  and  ancillary  responses generated  according  to  standard
pipelines                         and                        selection
criteria\footnote{\url{http://heasarc.nasa.gov/docs/xte/recipes/pca_spectra.html#analysis}}.
We used  only events from PCU2  data, as this PCU  more completely and
uniformly  overlapped  with   the  \textit{Chandra}  observation,  and
limited our spectral analysis to the top-layer events as these provide
the best signal-to-noise  spectra in the 3.5--25.0  keV energy range,
where the  response matrix is  best calibrated; a systematic  error of
1\%  was added  in quadrature  to  the statistical  error.  For  light
curves and hardness  ratios we exploited the  broader range 2.0--30.0
keV.  The details of the $RXTE$ observations used for the analysis are
summarized in Table~\ref{log_rxte}.

\begin{table*}
\centering      
\begin{tabular}{l c c c c c }
\hline\hline             
ObsId                     & TSTART               & Exposure & Rate    \\
                          & UTC                   & s        &  cts s$^{-1}$     \\
\hline
95338-01-01-04 &          2010-08-01   01:25:34  &  \phantom{0}752     &  715.6$\pm$1.0  \\
95338-01-01-03 & \phantom{2010-08-01}  03:00:46  &   1104     &  737.5$\pm$0.8  \\
95338-01-01-02 & \phantom{2010-08-01}  04:35:26  &   1328     &  780.2$\pm$0.8  \\
95338-01-01-01 & \phantom{2010-08-01}  06:09:34  &   2000     &  630.0$\pm$0.6  \\
95338-01-01-05 & \phantom{2010-08-01}  07:43:42  &   3152     &  594.4$\pm$0.4  \\
\hline
\hline                               
\end{tabular}
\caption{Log of the \textit{RXTE}/PCU observations. 
Net count rates and exposures for PCU2 top-layer spectra
are given. Start time (TSTART) is in UTC.}
\label{log_rxte}
\end{table*}

\section{Light curves and time-selected spectral analysis}

\begin{figure}
\centering{}
\includegraphics[width=\columnwidth, height=\columnwidth, angle=-90]{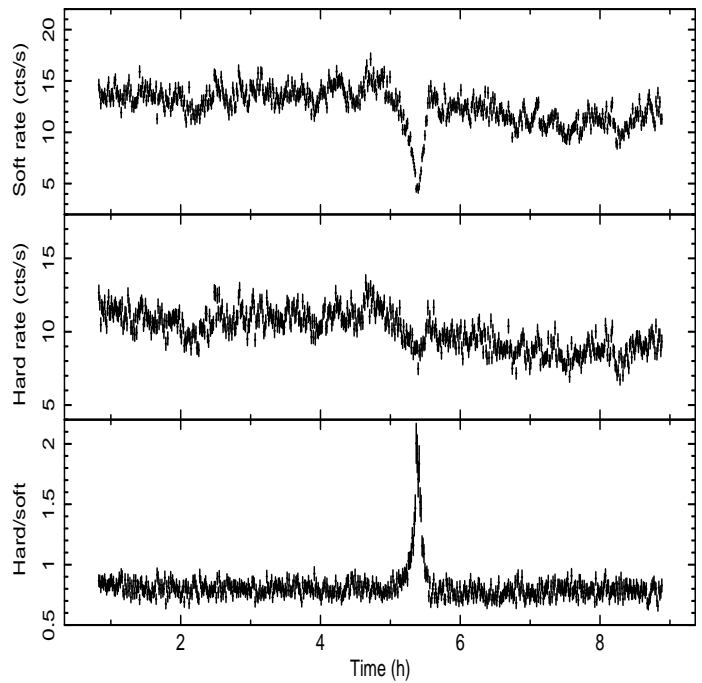}
  \caption{\textit{Chandra}/HEG light curve  and hardness ratio.  Time
    is  in  hours since  the  start  of  the day  2010-08-01  00:32:37
    UTC. The  soft rate refers to  the 1.0--4.0 keV range,  while the
    hard  rate refers  to the  4.0--10.0 keV  range. Bin  time is  50
    s. Re-adapted from \citet{iaria14}.}
\label{fig1}
\end{figure}

In Fig.~\ref{fig1}  we show  the \textit{Chandra}  light curve  in the
1.0--10.0  keV  range,  extracted  from the  ACIS-S  HEG  first-order
diffraction arm  and the hardness  ratio defined from  the (4.0--10.0
keV)/(1.0--4.0 keV) count  ratio.  The dip is  clearly chromatic, the
hardening is  smooth, and the  center of  the dip falls  at 2010-08-01
05:23:22.848  UTC;  the  dip,  which  is  neatly  resolved,  is  quite
symmetrical   with  ingress   and   egress  times   that  are   nearly
coincidental,  although  the  smooth   ingress/egress  times  and  the
correlated variation of  the persistent emission does not  allow us to
strictly define its  temporal duration.  Fitting the dip  shape with a
Gaussian in  a local (750 s  wide) neighborhood of the  dip center, we
derived a  full width  at half  maximum of $\sim$  450 s.   Before and
after  the dip,  the spectrum  shows  an overall  smooth and  moderate
softening.  In the pre-dip part  of the observation, the average count
rate   and  hardness   ratio   are  24.4$\pm$0.1   cts  s$^{-1}$   and
0.798$\pm$0.003,  in  the post-dip  part  these  are 20.2$\pm$0.1  cts
s$^{-1}$ and 0.773$\pm$0.003.

\begin{figure}
\centering{}
\includegraphics[height=\columnwidth, width=\columnwidth,  angle=-90]{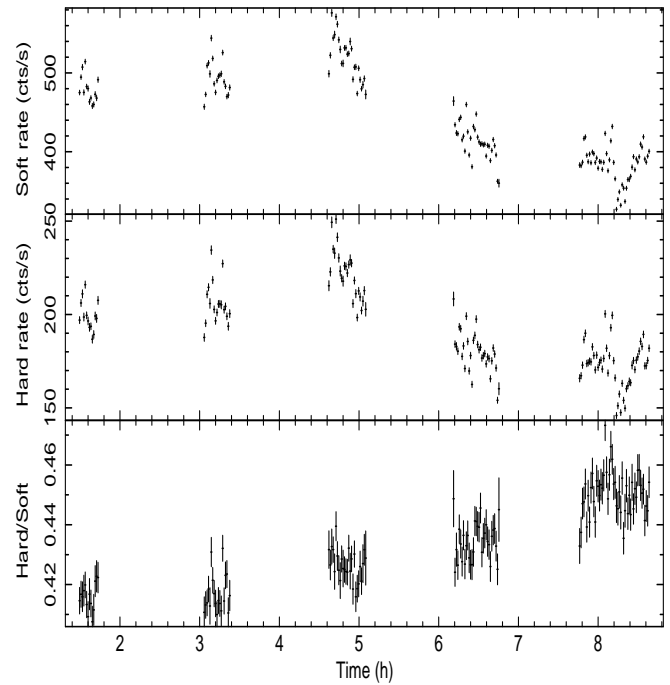}
\caption{Light  curves   and  hardness   ratio  for   the  $RXTE$/PCU2
  observations. Soft and  hard rates computed in the 2--6  keV and in
  the  6--10  keV  range,  respectively.   Time  is  in  hours  since
  2010-08-01 00:00 UTC.}
\label{rxte_hardness}
\end{figure}

In  Fig.~\ref{rxte_hardness}, we  show  the PCU2  count  rates in  the
selected  energy   band  2--6  and   6--10  keV  and   the  hardness
ratio. While there is modest variability in the first three pointings,
a significant  increase in the  hardness is  evident in the  last two.
But  the  dip  event,  observed  in  the  continuous  \textit{Chandra}
pointing, was unfortunately missed by $RXTE$.

On the basis  of these hardness ratios, keeping the  number of spectra
reasonably within the need to  have sufficient statistics to constrain
the  main  spectral parameters,  we  created  the good-time  intervals
(Table~\ref{time_intervalsxx})  and then  extracted the  corresponding
energy spectra.   \textit{RXTE}/PCA observations were summed  with the
\texttt{mathpha} tool, after we verified their spectral consistency.

\begin{figure}
\centering
\includegraphics[height=\columnwidth, width=0.8\columnwidth,  angle=-90]{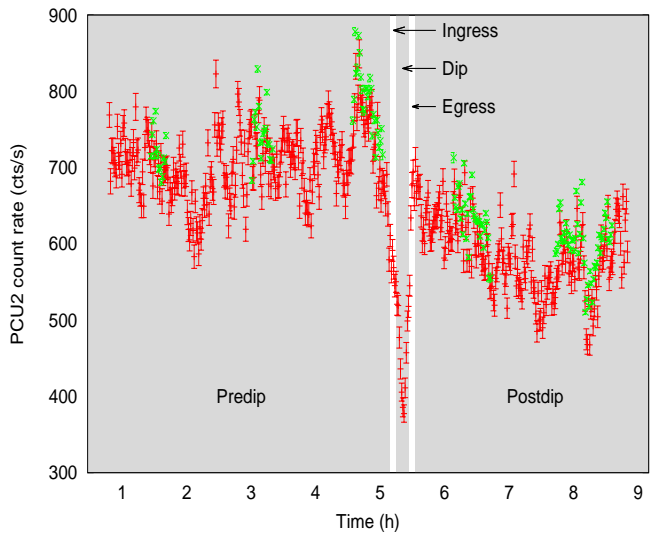}
\caption{2--30  keV PCU2  top  layer count  rate  (green points)  and
  1--10 keV \textit{Chandra}/HEG rate (red points).  \textit{Chandra}
  rate is multiplied by a factor of 24 to approximately match the PCU2
  rate. Bin  time is 64s.   Alternated shaded  areas are based  on the
  time-selections  of Table~\ref{time_intervalsxx}.  Time is  in hours
  since 2010-08-01 00:00 UTC}.
\label{chadrarxte_lc}
\end{figure}

In  Fig.~\ref{chadrarxte_lc} we  show  the  PCU2 pointed  observations
overimposed  to the  \textit{Chandra}  light curve  (the  HEG rate  is
multiplied by a factor of 24 to visually match the PCU2 data), and the
time selections through alternating white and gray overlays.  Two main
longer  intervals  identify  the  pre-dip and  post-dip  spectra;  one
interval is  centred around the  dip bottom and two  shorter intervals
trace the ingress  and egress passages (we did not  use a specific cut
on  the count  rate to  define these  intervals) for  a total  of five
time-selected spectra.

\begin{table}
\centering
\caption{Time-selected   intervals  used   for  the spectral analysis  of
  \textit{Chandra} and \textit{RXTE} observations.} 
\begin{tabular}{llll}
\hline \hline
Spectrum         & $T_{start}-T_{stop}$ & HEG rate         & RXTE datasets \\
                 & s          & cts s$^{-1}$     & \\
\hline
Pre-dip          &  \phantom{00}995--16734       &  24.43$\pm$0.04     & 04/03/02   \\
Ingress          &  16734--17084     &  19.4\phantom{0}$\pm$0.6     &  NONE   \\
Dip              &  17084--17784     &  15.2\phantom{0}$\pm$0.3     &  NONE   \\
Egress           &  17784--18134     &  22.0\phantom{0}$\pm$0.6              &  NONE  \\
Post-dip         &  18134--30065     &  20.30$\pm$0.05     &  01/05  \\
\hline \hline
\end{tabular}
\tablefoot{Start and stop times are relative  to the start time of the
  \textit{Chandra}  observation corresponding  to 2010-08-01  00:32:37
  UTC. The HEG rate is  the summed (positive and negative) first-order
  1.0--10  keV rate  (uncorrected for  pile-up).The \textit{RXTE}/PCA
  spectra used to cover the broad-band  spectrum are listed in the last
  column.}
\label{time_intervalsxx}
\end{table}

\section*{Spectral analysis}
\subsection{Pile-up treatment}
Because  of  the  intense  source flux  the  \textit{Chandra}  grating
spectra are moderately  affected by pile-up. To assess  its impact, we
used    in   the    spectral    fits    the   convolution    component
\texttt{simple\_gpile2} developed by \citet{hanke09} from the original
code in \citet{nowak08}.   The main effect of  pile-up for first-order
spectra is to reduce the effective  count rate in each detector pixel.
The \texttt{simple\_gpile2} uses the  free fitting parameter $\gamma$,
which corrects the detector count rate according to the equation
\begin{equation}
C' (\lambda) = C (\lambda) \cdot exp(- \gamma \cdot C_{tot}(\lambda)),
\end{equation}
where  $C (\lambda)$  is the  observed detector  count rate,  $C_{tot}
(\lambda)$ is  the count rate  computed according to the  total source
fluxes (summed over all orders), and $\gamma$ is expressed in units of
s \AA/cts.   The $\gamma$ parameter is  free to vary for  each HEG/MEG
dataset, with an expected value that is the function of the time frame
and detector wavelength accuracy, as  expressed by the simple relation
$\gamma_0 = 3 \Delta \lambda  \times t_{frame}$ where $\Delta \lambda$
corresponds to 5.5  m\AA~for the HEG grating arm, and  11 m\AA~for the
MEG arm, and $t_{frame}$ = 1.24  s for this observation.  The factor 3
takes  into  account that  the  detection  cell  is constituted  by  a
three-pixel array.

\begin{figure}
\centering
\includegraphics[height=\columnwidth, width=0.8\columnwidth,  angle=-90]{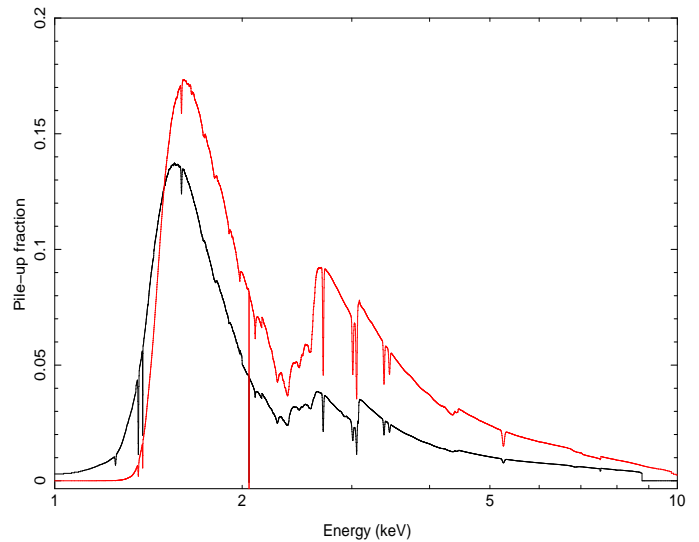}
\caption{Pile-up fraction as a function of energy for the $Chandra$/HETGS data. 
Black line HEG+1 data, red line MEG+1 data. Negative orders show a very similar dependence.}
\label{pileup_fraction}
\end{figure}

In  Fig.\ref{pileup_fraction},  we  show  the pile-up  fraction  as  a
function of  energy according to  the best-fitting model  discussed in
Section~\ref{broadband}.  We  found that  the $\gamma$  parameters did
not  strongly  depend  on  the  continuum  choice  and  showed  little
variation  for the  time-selected spectra.   The pile-up  fraction, as
expected,  reached higher  values  for the  MEG  spectra (peak  values
$\gtrsim$ 15\% between 1.54 keV and 1.74 keV) than for the HEG spectra
(peak values $\gtrsim$ 10\% between 1.4 keV and 1.8 keV).  The pile-up
fraction is negligible in the K$\alpha$ iron range.

\subsection*{Absorption line detections in the non-dip $Chandra$ data}  \label{linedetections}

The $Chandra$  X-ray spectrum  is rich  in local  absorption features,
similarly   to   the    \textit{Chandra}   observation   analyzed   in
\citet{ueda04}.  In  this  section,  we focus  on  the  detection  and
identification  of these  local  features, while  the  details on  the
continuum model  are addressed in  the next section.  To  increase the
signal-to-noise ratio and assess the detection level of these features
we  considered  the time-averaged  spectrum,  filtering  out only  the
interval with the  dip. We considered HEG$\pm$1 and  MEG$\pm$1 data as
independent datasets,  adopting a  continuum model consisting  of soft
disk emission  and thermal  Comptonized component  (the choice  of the
exact continuum  has marginal influence  on the parameter  values) and
locally searched  for narrow  absorption features around  the expected
values for  the resonant  transitions H-like and  He-like of  the most
abundant  elements,  fitting  the  line profiles  with  Gaussians.  In
Table~\ref{gaussians}  we  report  the detected  lines  (normalization
values  not compatible  with  0  at 2.7  $\sigma$)  associated to  the
element transitions, the measured shift with respect to the laboratory
rest-frame and  the widths. Two lines  close to the Si  K$\alpha$ edge
are possibly  not related to the  warm medium but to  X-ray absorption
fine   structures    (XAFS)   close    to   the    Si-edge   structure
\citep{ueda05}.   Best-fitting  values   for   the  pile-up   $\gamma$
correction  factors  applied for  each  grating  arm are  reported  in
Table~\ref{pileupgamma}.

\begin{table*}
\caption{Absorption  line  detections for  $Chandra$  non-dip
  spectrum.  An  absorbed  ($\texttt{tbvarabs}$)  continuum  model  of
  thermal disk emission and thermal Comptonization is used.}  \center
\begin{tabular}{lll lll ll}
\hline \hline
Ion &Transition\tablefootmark{a} & E$_{lab}$  & E$_{obs}$ & Shift & Width & Flux\tablefootmark{b} & EQW\tablefootmark{c}   \\
    &                            & eV     & eV          & km/s       & eV  & (10$^{-4}$)    &  eV       \\
\hline \hline
\ion{Mg}{xii}  & $1s-2p$    &1472.3 & 1474.0$_{-1.5}^{+1.0}$ & 350$_{-300}^{+200}$ & 1.5$\pm$1.5 & 6.5$\pm$3 & 0.8$\pm$0.3 \\
\ion{Al}{xiii} & $1s-2p$    &1728.6 & 1727.2$\pm$2.6         & -240$\pm$450        & 1.6$_{-1.6}^{+4}$ & 5$\pm$2.5 &  0.6$\pm$0.3\\
\ion{Mg}{xii}  & $1s-3p$    &1744.7 & 1745.4$\pm$1.2         & 140$\pm$200         & 1.0$_{-1.0}^{+1.6}$ & 6.0$\pm$2.0 & 0.9$\pm$0.3\\ 
XAFS           & &          & 1847.7$\pm$0.7         & & 2.2$\pm$0.8    & 11.6$\pm$2.4    & 2.0$\pm$0.4      \\
XAFS           & &          & 1863.7$\pm$1.0         & & 2.4$\pm$1.4    & 9$\pm$2         & 1.8$\pm$0.3 \\
\ion{Si}{xiv}  &$1s-2p$     &2005.5 & 2007.4$\pm$0.4         & 280$\pm$60          & 2.6$\pm$0.6  & 14.4$_{-1.4}^{+1.7}$ & 2.7$\pm$0.5 \\
\ion{S}{xvi}   &$1s-2p$     &2621.7 & 2623.4$^{+0.6}_{-1.2}$ & 200$^{+70}_{-140}$  & 0$^{+4}$ & 6.0$\pm$1.4  & 1.2$\pm$0.4 \\
\ion{Ca}{xx}   &$1s-2p$     &4105.0 & 4118$\pm$8             & 950$\pm$600         & 13$\pm$10 & 6.0$\pm$2.5  & 2.4$\pm$1.1 \\
\ion{Fe}{xxv}  &$1s^2-1s2p$ &6700.4 & 6706$\pm$5             & 250$\pm$220         & 8$\pm$8  & 9.6$\pm$2   & 12$\pm$3  \\
\ion{Fe}{xxvi} &$1s-2p$     &6966.2 & 6978$\pm$3             & 500$\pm$130         & 20.4$\pm$5 & 25.2$\pm$3 & 36$\pm$5 \\ 
\ion{Fe}{xxvi} &$1s-3p$     &8250.2 & 8273$\pm$20            & 840$\pm$730         & 25$\pm$25  & 10$\pm$5 & 22$\pm$12 \\
\hline 
\hline
\end{tabular}
\tablefoot{
\tablefoottext{a}{Rest-frame energies from \citet{verner96}.}
\tablefoottext{b}{Total area of the Gaussian (absolute value), in units of photons/cm$^{-2}$/s.}
\tablefoottext{c}{Line equivalent width.}
}
\label{gaussians}
\end{table*}

Inspecting the results from Gaussian  fitting of the absorption lines,
we noted that the lines are mostly produced by resonant transitions of
H-like  ions  with  a  common  blue-shift  (weighted  average  490  km
s$^{-1}$).  Under the assumption that all these features could then be
described  by  only one  photoionized  medium,  we adopted  a  tabular
spectral      model     derived      from     the      XSTAR     code,
\texttt{warmabs}\footnote{\url{http://heasarc.gsfc.nasa.gov/xstar/docs/html/node99.html}}
to self-consistently fit  all these features.  We set  as free fitting
parameters  the  log of  the  ionization  factor $\xi$,  the  relative
hydrogen column  density (in  units of  log(N$_{\textrm H}/10^{22}$)),
the blue-shift  of all the  lines (using the $z$  red-shift parameters
and allowing for negative values)  and the turbulent broadening (in km
s$^{-1}$).   We  adopted as  electron  density  a value  of  10$^{12}$
cm$^{-3}$,  although we  note that  this  choice has  only a  marginal
impact on the best-fitting values of all the other parameters. We used
the  default value  of the  table model  for the  irradiating flux,  a
power-law with spectral  index 2. This is a good  approximation of the
1--10  keV  spectrum of  the  source,  since  fitting with  a  simple
absorbed  power-law  would result  in  a  photon-index of  2.07.   The
best-fitting  values  for  the   \texttt{warmabs}  component  and  the
continuum model  adopted are discussed  and reported in  the following
section  (last   column  of   Table  \ref{pre-post-dip}),   while  the
best-fitting  values  of  the  pile-up  $\gamma$  parameters  for  the
different grating arms are reported in Table \ref{pileupgamma}.

\begin{table}
\centering
\begin{tabular}{lll}
\hline \hline
&Gaussian Fit & Warmabs Fit \\ 
\hline
Grating arm & \multicolumn{2}{c}{$\gamma$  \small{(10$^{-2}$ s $\cdot$ \AA/cts)}} \\
HEG+1       & 2.7$\pm$0.3 & 3.0$\pm$0.3 \\
HEG-1       & 2.9$\pm$0.3 & 3.1$\pm$0.3\\
MEG+1       & 3.2$\pm$0.3 & 3.4$\pm$0.3 \\
MEG-1       & 3.2$\pm$0.3 & 3.4$\pm$0.3\\
\hline \hline
\end{tabular}
\caption{Best-fitting    values    of     the    averaged    $Chandra$
non-dip spectrum  for  the  $\gamma$ pile-up parameters.}
\label{pileupgamma}
\end{table}

\begin{figure*}
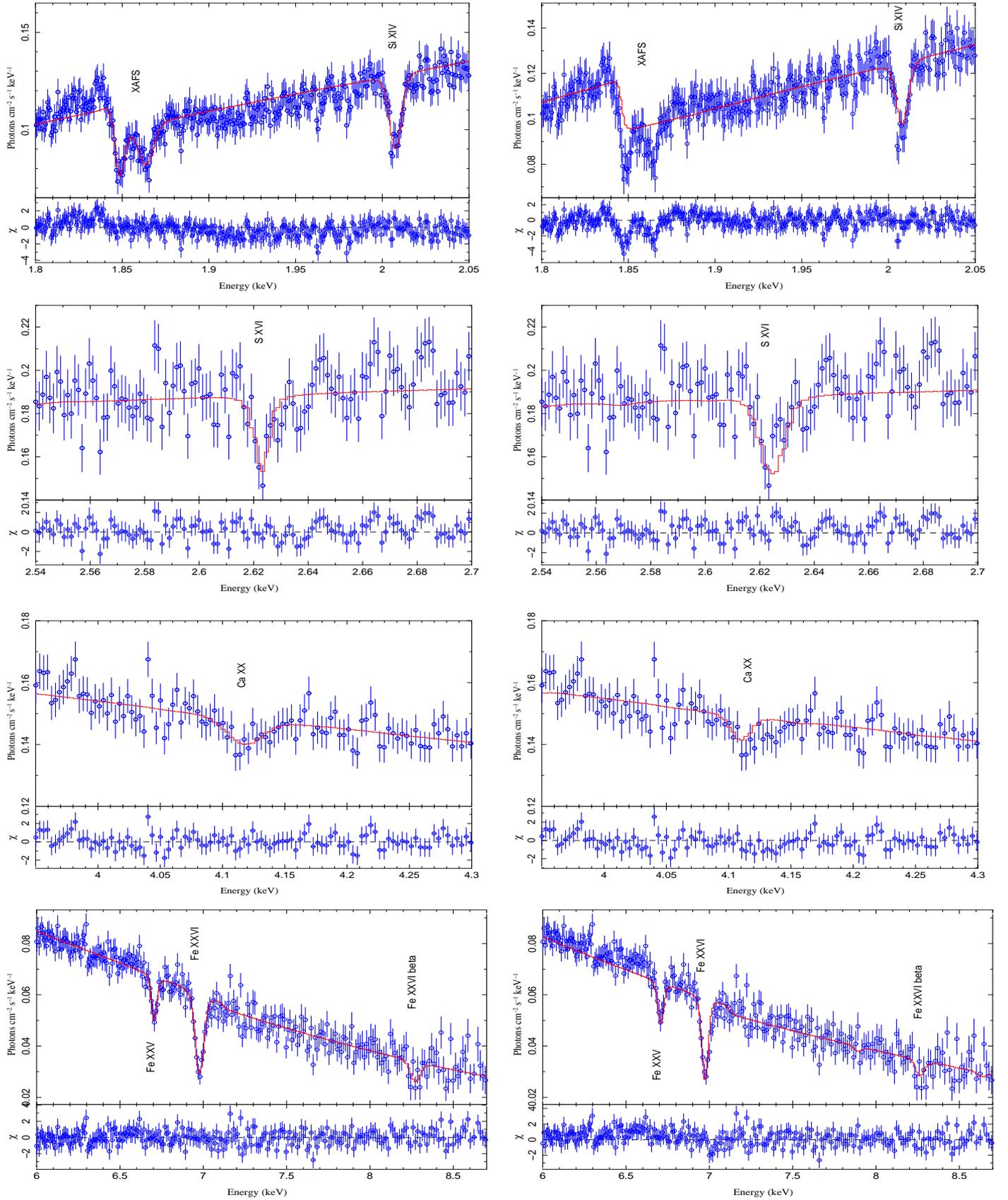

\centering
\begin{tabular}{cc}
\includegraphics[height=\columnwidth, width=0.6\columnwidth,  angle=-90]{f05.ps}& 
\includegraphics[height=\columnwidth, width=0.6\columnwidth,  angle=-90]{f06.ps}\\

\includegraphics[height=\columnwidth, width=0.6\columnwidth,  angle=-90]{f07.ps}& 
\includegraphics[height=\columnwidth, width=0.6\columnwidth,  angle=-90]{f08.ps}\\

\includegraphics[height=\columnwidth, width=0.6\columnwidth,  angle=-90]{f09.ps} &
\includegraphics[height=\columnwidth, width=0.6\columnwidth,  angle=-90]{f10.ps}\\

\includegraphics[height=\columnwidth, width=0.6\columnwidth,  angle=-90]{f11.ps}& 
\includegraphics[height=\columnwidth, width=0.6\columnwidth,  angle=-90]{f12.ps}\\
\end{tabular}
\caption{Data, best-fit  model, and residuals for  detected absorption
  local  features of  the  \textit{Chandra}  non-dip average  spectrum
  (HEG+MEG data combined  for clarity), using a continuum  model of an
  absorbed  accretion disk  and thermal  Comptonized component.   Left
  panels:  lines fitted  with Gaussian  profiles; right  panels: lines
  fitted with the XSTAR warm-absorber model.}
\label{comparison}
\end{figure*}

In Fig.~\ref{comparison}, we  compare model and residuals  for the two
approaches (local  Gaussians and \texttt{warmabs} model)  for the most
interesting  energy  ranges.   The  strongest  detected  features  are
satisfactorily fitted  with the  \texttt{warmabs} component,  with the
exception  of the  local features  around the  K$\alpha$ Si-edge  (see
upper  right  panel  of  Fig.\ref{comparison}).   For  the  ionization
parameter   derived  from   the  best-fitting   model,  the   possible
contribution of the  Ly$\alpha$ transition of He-like Si  at 1.865 keV
is negligible, and  therefore we conclude that the  most likely origin
of  these  features is  the  presence  of  XAFS. The  overall  reduced
$\chi^2$  ($\chi^2_{red}$) of  the model  adopting local  Gaussians is
1.074 (7793  degrees of freedom,  [d.o.f.]), while the model  with the
\texttt{warmabs} gave  1.102 (7818 d.o.f.). Adding  two Gaussians with
energies   and   widths   corresponding   to  the   XAFS   values   of
Table~\ref{gaussians} gave a satisfactory account of the residuals and
a $\chi^2_{red}$ of  1.080, for 7812 d.o.f.   Although the statistical
difference in  the two fits  is significant, the matching  between the
position,  widths, and  broadening  of the  highly ionized  absorption
lines in  the two approaches is  remarkable, and we conclude  that the
key   physical  characteristics   of  the   absorption  features   are
satisfactorily accounted for by a uniformly photoionized wind.

\subsection{Broad-band continuum model} \label{broadband}
To fit the broad-band continuum, we modeled the interstellar absorption
with  the \textit{tbvarabs}  spectral  component,  with the  abundance
table of \citet{asplund09} and  the cross section of \citet{verner96}.
We forced the Mg, S and Fe  abundance to assume the same value as that
of the Si abundance and left  this parameter free to vary with respect
to  all the  other  elements (which  effectively  implies a  different
abundance  ratio of  the high-Z  elements  with respect  to the  low-Z
ones),  according  to the  possible  overabundance  in heavy-Z  metals
reported in the study of  \citet{ueda05}.  In addition to the resonant
absorption  lines,  the warm  absorber  is  also responsible  for  the
continuum  optically thin  Compton  scattering.  To  account for  this
effect,  we  multiplied  the  model   by  an  optically  thin  Compton
scattering component (\texttt{cabs} in Xspec), tying the \texttt{cabs}
electron density to 1.3 times  the value of the \texttt{warmbs} column
density, because  we assumed  a metallicity  ($Y+Z$=0.3) for  the warm
absorbing  medium.   We used  a  multiplication  constant between  the
$Chandra$ and  \textit{RXTE}/PCA models to take  into account possible
calibration differences in the observed flux.

To  determine the  continuum emission  we used  the $Chandra$/HEG  and
\textit{RXTE}/PCA spectra; we tried  two different spectral broad-band
models, but  we limited our  comparison to the pre-dip  spectrum.  The
first model  follows the so-called  extended coronal model,  which has
been extensively  adopted for dipping  sources and more  recently also
for     bright    Z-sources     \citep{balucinska11,church12}.    This
model\footnote{\url{constant*simple_gpile2*tbvarabs*cabs*
    warmabs*(cutoffpl +  blackbody + gaussian)} in  Xspec language} is
composed of a cut-off power-law to take into account emission from the
harder  coronal environment,  a thermal  black-body component  to take
into account  direct emission from the  NS, and a broad  iron emission
line,  with  an upper  limit  to  the width  at  0.5  keV, to  prevent
unphysical  continuum distortions,  and a  line energy  constrained to
take values between 6.4 keV and 7 keV.
 
This  model gave  a  rather flat  value for  the  photon-index of  the
cut-off power-law,  however, with a best-fitting  value of 1.1$\pm$0.3
and a  cut-off energy of  4.4$\pm$0.4 keV,  while the radius  from the
thermal black-body component  is $\sim$ 4 km, and  might be compatible
with a narrow  equatorial strip of the NS surface  at a temperature of
1.00$\pm$0.05 keV.  The  reduced $\chi^2$ for this fit  was 1.06 (2416
d.o.f) and no other significant  residual pattern was evident.  If the
index of  the power-law is  associated to a Comptonized  spectrum, the
low index  value becomes difficult  to interpret, because  the cut-off
energy is  $\sim$ 4.4  keV. For  a spectral  index $\alpha$  $\sim$ 0,
there  is no  physical  solution that  can be  converted  in terms  of
physical parameters \citep{pozdnyakov83}.

We then  adopted a model  according to the  so-called \textit{Eastern}
decomposition,  composed of  a  multitemperature accretion-disk  model
with  a  zero-torque  boundary condition  \citep[\texttt{ezdiskbb}  in
  Xspec,][where  we assumed  a distance  of 7  kpc, a  color-effective
  temperature  ratio of  1.7 and  an inclination  angle of  65 deg  to
  calculate the  disk inner  radius]{zimmerman05} to model  the softer
energies  and  a Comptonization  model  to  fit the  high-energy  band
\citep[\texttt{comptt}    in     Xspec,][]{titarchuk94}.     In    the
Comptonization component  we assume spherical geometry.   The spectral
parameters that  determine the shape  of the Comptonized  spectrum are
the  soft seed-photon  temperature  ($kT_s$), the  temperature of  the
electron  cloud ($kT_e$),  and the  optical depth  ($\tau$). For  this
model
\footnote{\url{constant*simple_gpile2*
    tbvarabs*cabs*warmabs*(ezdiskbb+comptt+gaussian)}  in  Xspec},  we
found a strong correlation between  the disk temperature, and the soft
seed-photon temperature.  This is expected  since for X-ray sources at
high inclination angle the observed disk emission, which scales as the
cosine of the angle, becomes weaker with respect to the more isotropic
boundary layer emission.

\begin{figure}
\centering
\includegraphics[height=\columnwidth, width=\columnwidth,  angle=-90]{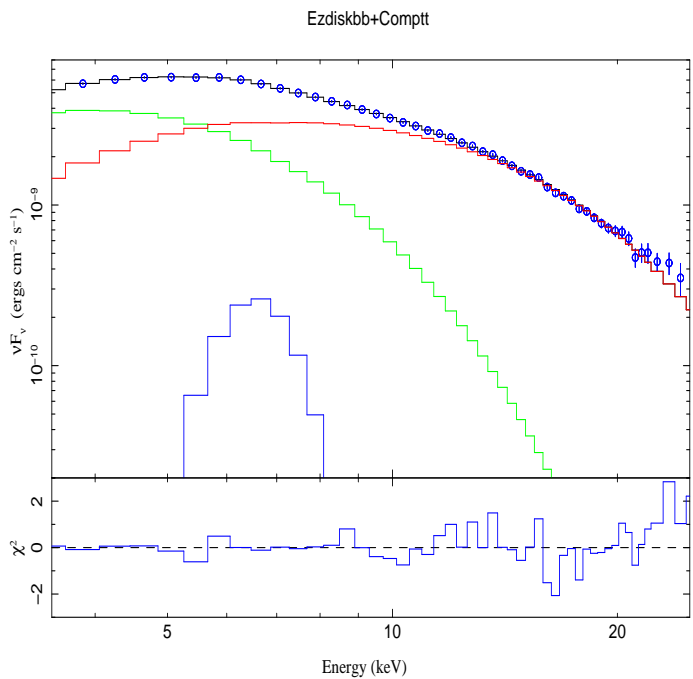} \\
\caption{$RXTE$/PCA pre-dip (3.5--25 keV range) unfolded spectrum, with contributions from spectral 
components and residuals, adopting the Eastern decomposition: green, red, and blue lines show contribution from
accretion disk, Comptonized component, and Gaussian line, respectively.}
\label{fig_comparison}
\end{figure}

However,  for similar  accretion  rates, in  low-inclination and  less
absorbed  LMXB, where  disk  emission is  clearly  separated, the  two
temperatures are usually  found to have very similar  values, with the
disk    temperature   typically    20\%-30\%    lower   than    kT$_s$
\citep{disalvo02}, because the Comptonization is thought to act mainly
on the hotter  photons produced at the surface of  the NS.  Because of
the overlapping range of values for  the two parameters, we decided to
force them to assume the same value  to make the fit more stable.  For
this  model  we  derived  an  inner  radius  for  the  accretion  disk
compatible with a  truncation radius at the boundary  layer ($\sim$ 15
km), which for  high-accretion rates can extend to  a distance similar
to the  radius of the  NS \citep{dai10}. The highest  disk temperature
(set  equal to  the photon-seed  temperature) value  is $\sim$  1 keV,
while  the high  optical depth  ($\tau$ $\sim$  8.5) and  low electron
temperature ($\sim$ 3 keV) are consistent with the general softness of
the spectrum.  A moderately broad  iron line in emission  is required.
The line  equivalent width is  $\sim$ 50 eV, significantly  lower than
the    values    reported    in    \textit{XMM-Newton}    observations
\citep{diaztrigo12}.  We note,  however,  that the  line detection  is
mostly  driven  by residuals  of  the  \textit{RXTE}/PCA, while  using
$Chandra$/HEG data alone the detection would be less constraining.  We
show  the broad-band  $RXTE$/PCA  unfolded spectra  together with  the
contributions   from  the   additive  components   and  residuals   in
Fig.~\ref{fig_comparison}.

Because  the best-fitting  parameters  of the  model  and the  general
spectral  decomposition agree  with the  results of  other LMXBs  that
accrete  at  similar  rates,  we  hold  this  continuum  model  to  be
adequately solid and adopted it  for the other time-selected intervals
and  to obtain  the best-fitting  averaged continuum  emission of  the
whole out-of-dip $Chandra$  observation.  In Table \ref{pre-post-dip},
we report  the best-fitting values  and associated errors for  the two
time-selected  pre-dip  and  post-dip  spectra  and  for  the  average
$Chandra$ observation.  The most  significant changes from  pre-dip to
post-dip parameters are on the fluxes  of the two components (a change
of $\sim$  20\%, while the  statistical error  on a single  measure is
$\lesssim$ 3\%), and a change in  the optical depth of the Comptonized
component, while  the other  spectral parameters are  found consistent
within the statistical uncertainties. The average out-of-dip $Chandra$
observation presents a higher value  for the N$_{\textrm H}$ parameter
than  that of  the RXTE+HEG  fit, but  the error  on the  parameter is
underestimated because we kept the  Mg/Si/S/Fe abundance frozen to the
reference  value of  1.3,  as  the two  parameters  were  found to  be
strongly correlated.

\begin{table*}
\caption{Fitting results for the time-selected out-of-dip spectra and for the 
$Chandra$ average spectrum.} \label{pre-post-dip}
\center
\begin{tabular}{l|l|l|l}
\hline \hline
                                                          & \multicolumn{2}{c}{PCA+HEG} & $Chandra$ (HEG+MEG)               \\
Parameters                                                &  Pre-dip    &   Post-dip  &  Average out-of-dip  \\                
\hline

N$_{\textrm H}$ (10$^{22}$ cm$^{-2}$)                     & 3.7$\pm$0.2   & 3.8$\pm$0.2 & 4.10$\pm$0.06\\
\vspace{2mm}
Mg/SI/S/Fe abundance                                      & 1.3$\pm$0.2   &  1.4$\pm$0.3 &  1.3\tablefootmark{$\dag$}  \\

Disk kT (keV) /Comptt  kT$_s$ (keV)        & 0.97$\pm$0.02 & 0.98$\pm$0.04 & 1.04$\pm$0.05     \\
Disk inner radius\tablefootmark{a}   (km)  & 15.2$\pm$2.7  & 19$\pm$3      & 18.8$\pm$2.8 \\
\vspace{2mm}
Disk flux\tablefootmark{b} (10$^{-9}$ erg cm$^{-2}$ s$^{-1}$)  & 2.8 &  4.0 &  5.5 \\

Comptt kT$_{e}$ (keV)                               & 3.07$\pm$0.09 & 3.10$\pm$0.07&3.6$\pm$0.4  \\
Comptt $\tau$                                       & 8.5$\pm$0.3   & 9.6$\pm$0.3  &8.5\tablefootmark{$\dag$}\\
\vspace{2mm}
Comptt flux\tablefootmark{b} (10$^{-9}$ erg cm$^{-2}$ s$^{-1}$)   &  13.4  &   9.2 & 10.5  \\

E$_{broad}$ (keV)                      & 6.6$\pm$0.1   &  6.5$_{-0.1}^{+0.2}$ & 6.4$^{+0.15}$    \\
Width (keV)                            & 0.2$\pm$0.15  &  0.5$_{-0.2}$        & 0.5$_{-0.1}$ \\
\vspace{2mm}
EQW   (eV)                             & 50$_{-30}^{+40}$& 90$\pm$50  & 45$\pm$20\\ 

CABS\tablefootmark{d} (10$^{22}$ cm$^{-2}$)             & 47$\pm$5       & 36$\pm$8      & 36$\pm$3  \\
Warmabs N$_{\textrm H,wa}$ (10$^{22}$ cm$^{-2}$)          & 36$\pm$4       & 28$\pm$6      & 28$\pm$2  \\
Warmabs log($\xi$)                     & 3.60$\pm$0.1   & 3.66$\pm$0.10 & 3.65$\pm$0.05    \\
Warmabs broadening (km s$^{-1}$)       & 380$\pm$80     & 450$_{-150}^{+400}$ & 450$_{-100}^{+50}$\\ 
\vspace{2mm}
Warmabs blueshift (km s$^{-1}$)        & 420$\pm$90     & 500$\pm$130 & 420$\pm$60 \\

Constant (PCA/HEG)                    & 1.08$\pm$0.01   & 1.08$\pm$0.01 & \\
\vspace{2mm}
\vspace{1mm}
$\chi^2_{red}$ (d.o.f.)     & 1.06 (2416) & 1.08 (1998) &  1.08 (7812)  \\
\hline \hline
\end{tabular}
\tablefoot{
HEG+1, HEG-1 data combined in the 1.0--10.0 keV band. $RXTE$/PCA data in the 3.5--25 keV band.
Model \texttt{gpile*constant*tbvarabs*cabs*warmabs*(ezdiskbb+comptt+gau)}.\\
\tablefoottext{$\dag$}{Frozen parameter}
\tablefoottext{a}{Assuming a distance of 7 kpc, a hardening factor of 1.7, and an inclination angle of 65\degr.}
\tablefoottext{b}{Unabsorbed extrapolated flux in the 0.1--100 keV energy range.}
\tablefoottext{c}{This parameter is tied to be 1.3$\times$ the \texttt{warmabs} N$_{\textrm H,wa}$ component.}}
\end{table*}

\subsection*{Spectral changes during dipping}

To  study  the   spectral  changes  during  the  dip,   we  used  only
\textit{Chandra}  HEG  and   MEG  data,  as  there   was  no  strictly
contemporaneous \textit{RXTE}  observation of  source during  the dip.
Because of the short duration of the dip and the low effective area of
the $Chandra$/HETGS, no clear absorption feature was detected, so that
to  better constrain  spectral variations  we coarsely  re-binned each
grating arm data up to 100 counts per channel.  We assumed that during
the dip the continuum spectral shape is consistent with the out-of-dip
$Chandra$ average spectrum, because the dip falls approximately at the
center  of  the  observation.  We  therefore  kept  all  the  spectral
parameters frozen to  the best-fitting average values  of the spectrum
analyzed  in  Sect.  \ref{linedetections}  with  the  \texttt{warmabs}
component    (i.e.      $Chandra$    average    column     of    Table
\ref{pre-post-dip}).

\begin{figure*}
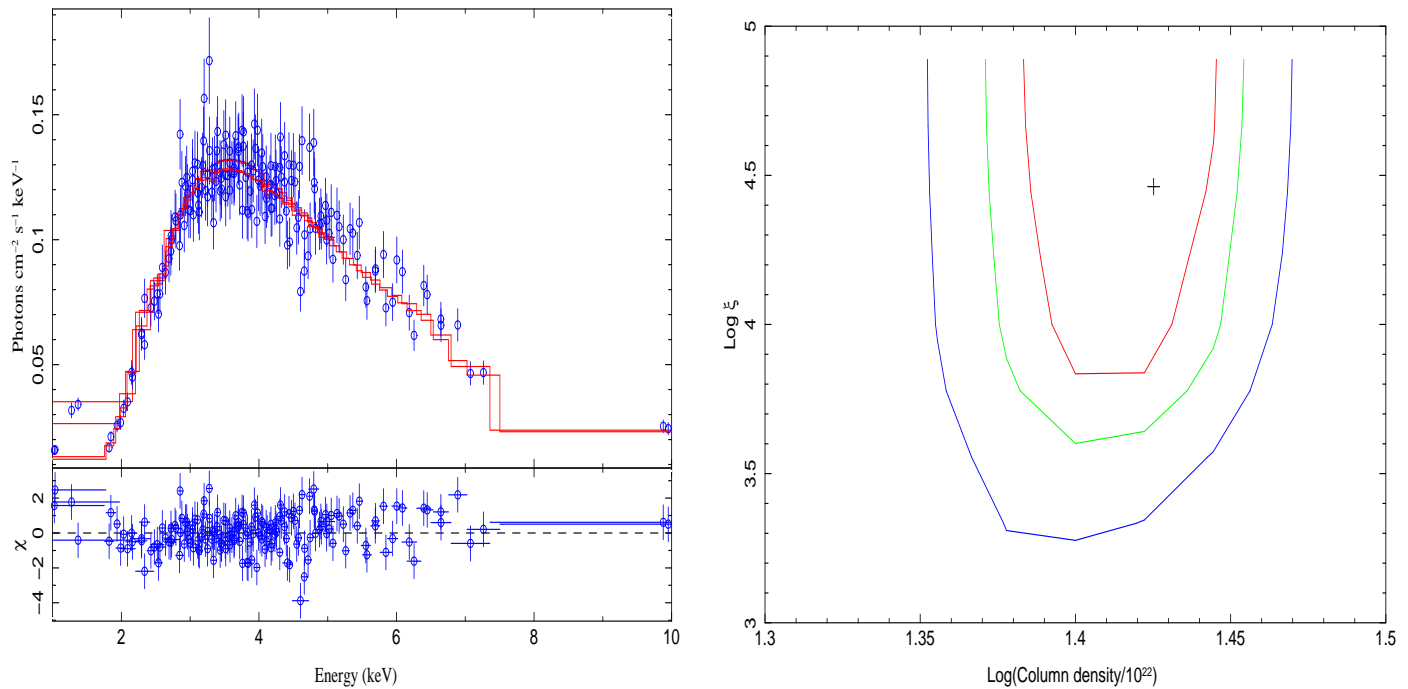

\centering
\begin{tabular}{cc}
\includegraphics[height=\columnwidth, width=\columnwidth,  angle=-90]{f14.ps} &
\includegraphics[height=\columnwidth, width=\columnwidth,  angle=-90]{f15.ps} \\
\end{tabular}
\caption{Left panel: data, unfolded best-fitting spectral model, and residuals for the time-selected
dip spectrum (HEG+MEG data combined). Right panel: contour plots for the logarithm of \texttt{warmabs} column
density (in units of 10$^{22}$ cm$^{-2}$) and log of ionization parameter. Red, green, and blue
contours indicate confidence levels at 68\%, 90\%, and 99\%, respectively.}
\label{dip_spectra}
\end{figure*}

We first  tried to model  the time-selected dip spectra  allowing only
for  a  variation  of  the  neutral  absorption  column,  keeping  the
parameters  of the  \texttt{warmabs} component  frozen to  the average
values.   A  multiplicative constant  before  the  model was  used  to
evaluate possible flux variations  between this time-selected spectrum
and  the averaged  one.  The  residual pattern  can be  satisfactorily
flattened, and we noted no  other significant residual. Intrinsic flux
variations with  respect to the  averaged spectrum  are only of  a few
percent, as the multiplicative constant is  very close to unity in all
examined  spectra  (see Table  \ref{dip_fits}).   We  also tested  the
possible presence of a neutral  partial covering effect.  The covering
fraction was  always compatible  with total  coverage, and  only lower
limits  could be  assessed,  without significant  improvements in  the
$\chi^2$ of the  fits.  The ingress and egress spectra  also show that
the  variation on  the cold  column density  is possibly  smooth, with
values  slightly  higher than  the  corresponding  values reported  in
Table~\ref{pre-post-dip}  for  the  pre-dip  and  post-dip  persistent
spectra.  This is  consistent with the hypothesis that  the smooth dip
profile is caused at first by an  increase and then a decrease of this
parameter alone.   In a second  round of  fits, we tested  whether the
spectral change is  compatible with a variation of  the column density
and  the  ionization state  of  the  \texttt{warmabs} component  alone
(keeping the neutral  N$_{\textrm H}$ fixed to the  average value). If
the multiplicative constant is kept  fixed to unity, no acceptable fit
is obtained ($\chi_{red}$ $>>$ 2),  whereas when we allowed a variable
constant,  we noted  a  statistically  acceptable fit  ($\chi^2_{red}$
1.08) for the  dip spectrum with a strong decrease  in both the column
densities (N$_{H,  wa}$ $\sim$8 $\times$ 10$^{22}$  cm$^{-2}$) and in
the  ionization state  (log($\xi$) $\sim$  -0.3), in  contrast with  a
general \textit{increase} in warm column  densities during dips in all
other dipping sources \citep{diaztrigo06}. However, the model constant
correspondingly  decreases  by  $\sim$20\%  in  correlation  with  the
decrease  of   the  \texttt{cabs}   value,  which   is  tied   to  the
\texttt{warmabs} column  density, so  that we  conclude that  this fit
artificially  reproduces the  former model,  without being  physically
consistent.  Finally, we evaluated whether  a change in the ionization
parameter and in  the absorption column of the warm  absorber could be
assessed  at  the  same  time,  but   we  found  marginal  or  no  fit
improvements  compared  with the  best-fitting  model  shown in  Table
\ref{dip_fits}.   For  the dip  spectrum,  we  found that  the  column
density of the warm absorber is  the same as in the averaged spectrum,
while for the  ionization parameter we found a  lower limit compatible
with the out-of-dip value (log($\xi$) =  3.6), as shown by the contour
plots of  the two  \texttt{warmabs} components in  the right  panel of
Fig. \ref{dip_spectra}.   Similar results  were also obtained  for the
ingress and egress spectra.

\begin{table*} \centering
\caption{Best-fitting spectral parameters for ingress, egress, and dip time-selected
spectra.} \label{dip_fits}
\begin{tabular}{lll l}
\hline \hline
Parameter (units) &  Ingress & Dip & Egress  \\                
$N_\textrm{H,cold}$ (10$^{22}$ cm$^{-2}$)      & 5.8$_{-0.2}^{+0.1}$  & 8.62$\pm$0.16 &  4.72$\pm$0.10\\
$C_{model}$                                    & 0.99$\pm$0.2         & 0.99$\pm$0.02 &  1.04$\pm$0.03\\
$\chi_{red}^2$ (d.o.f.)                        & 1.06 (123)           & 1.02 (178)    &  1.12 (142) \\
\hline
\end{tabular}
\tablefoot{The best-fitting model only has  the column density of the \texttt{tbvarabs} component
and a global model multiplicative constant ($C_{model}$) allowed to vary. All other parameters
are frozen to the average out-of-dip $Chandra$ spectrum (third column, Table \ref{pre-post-dip}).}
\end{table*}

In summary, our results indicate that the main driver of the dip event
is a rise in the column density  of a neutral (or very mildly ionized)
component,  without  requiring  any  change  of  the  \texttt{warmabs}
properties  with  respect to  the  out-of-dip  interval.  Keeping  the
neutral  column density  fixed at  the out-of-dip  averaged value,  no
physical or  statistically acceptable  solution is  found in  terms of
variation of the warm absorber properties.

\section{Discussion}

\subsection{Continuum formation and warm absorption}

We investigated the  1.0--25.0 keV spectrum of GX  13+1 by exploiting
the high-resolution \textit{Chandra}/HETGS spectrum, which has allowed
tight control  of the physical  characteristics of the  warm absorber,
and  \textit{RXTE}/PCA spectrum,  that  has allowed  to constrain  the
broad-band  continuum. Comparing  different spectral  models, we  found
that  the  parameters of  the  warm  absorber,  as expected,  are  not
strongly influenced  by the choice  of the underlying  continuum.  The
continuum  spectrum  is  extremely soft,  closely  resembling  typical
Z-sources spectra.  The most  physically plausible model is consistent
with a hot  accretion disk, whose apparent inner  radius is compatible
with the extension of the boundary layer for such high accretion rates
\citep{popham01},  and  an   optically  thick  thermal  Comptonization
component.  The total luminosity of the source is close to 50\% of the
Eddington limit for a canonical 1.4  M$_{\odot}$ NS, and it is similar
to  past  observations,  pointing  to a  certain  long-term  accretion
stability.   The  short-term  smooth   spectral  variability  that  we
observed  in  the  $\sim$30  ks  \textit{Chandra}  observation  mostly
reflects a  change in the  flux of  the components, while  the general
spectral shape remains unvaried.

Moreover,  energetically the  overall  emission  is probably  strongly
dominated by coronal power, and it  is unclear how the accretion power
can be efficiently transferred at very large radii, while the possible
power extractable from the accretion disk may be only a small fraction
of   the   energy   carried   by  the   disk   \citep[see   discussion
  in][]{uzdensky13}.

The  continuum is  absorbed  by  a warm  optically  thick plasma.  Its
characteristics are similar to what was shown in past \textit{Chandra}
observations,  and  we found  no  strong  variability in  the  present
observation, because  the best-fitting parameters between  the pre-dip
and  post-dip  spectra  are   all  consistent  within  the  associated
uncertainties.  Inspecting the  landscape of  the absorption  features
more closely, we found no evidence of the H-like transition of Mn, Cr,
and Ar reported  in \citet{ueda04}, but we  significantly detected the
Ly$\alpha$ transition of \ion{Al}{xiii} and Ly$\beta$ transitions from
\ion{Mg}{xii} and  \ion{Fe}{xxvi}, together with two  absorption lines
close to  the Si  K-$\alpha$ edge  related to  interstellar absorption
\citep{ueda05}.
 
The  self-consistent  model  \texttt{warmabs} yielded  a  satisfactory
representation of all the local features.  The average escape velocity
of the photoionized medium is $\sim$ 420$\pm$60 km s$^{-1}$; a similar
value was found in the  observation of \citet{ueda04}, while the lines
appear broadened  by the same order.  This is expected if  the wind is
thermally driven and  pushed by pressure gradients. In  this case, the
thermal motion of the ions would  be similar to the sound speed $c_s$,
that would be  similar to the escape velocity.   Roughly speaking, the
radius at which the wind is  launched should be larger than the radius
for  which the  escape  velocity is  the value  we  measured, so  that
R$_{wind}$ $>$ 2$\times$10$^{11}$ cm.  A  similar value is obtained by
a first-order calculation, taking  the ionization parameter derived by
the fit  ($\xi$ $\sim$  4000), a density  of 10$^{12}$  cm$^{-3}$, and
observed  luminosity (L$_{x}$  $\simeq$ 10$^{38}$  erg s$^{-1}$  for a
distance of  7 kpc), holding  a distance $r=\sqrt{L_x/(n  \cdot \xi)}$
$\sim$ 1.4 $\times$ 10$^{11}$ cm.

\subsection{Nature of the dip in the light curve}

A comprehensive study  on the dipping activity of the  source during a
time  span of  more than  14 years  was presented  in \citet{iaria14},
where evidence was given for a long-term modulation of dip passages at
an inferred  period of 24.5274(2)  d. The  phase of the  $Chandra$ dip
corresponds  to  the  zero-phase   of  the  periodic  modulation,  and
therefore this is the first spectroscopic study of the periodic dip in
GX 13+1.

The lack of past dip evidence in GX 13+1 may be attributed to the very
long  orbital period  of the  source, which  exceeds the  rest of  the
dipping class by almost two orders  of magnitude, and the possible low
duty-cycle of  the dips.  In  most dipping sources the  dip duty-cycle
may be as  large as half of  the entire orbital cycle  \citep[as in XB
  1323-619][]{balucinska99},   while,    for   longer-period   dippers
(P$_{orb}$ $>$ 1 day)  the duty cycle drops to a  few percentage as in
the case of  GRO J1655-40, and 4U 1639-47, or  Cir X-1, which implyies
that structure  that causes  the dip  does not  simply scale  with the
system dimensions. If the structure is identified with a bulge, formed
at the  impact point between the  accretion stream and the  outer disk
radius, then its physical dimensions  are probably determined by local
conditions. In  this case, the  duty cycle would inversely  scale with
the accretion-disk radius.

Indeed,  searching all  the  \textit{RXTE}  PCA pointed  observations,
\citep{iaria14} found  no clear  dip episode, the  dip spotted  in the
\textit{Chandra} observation  has a duration (FWHM)  of only $\sim$450
s, with very regular and smooth  ingress and egress times.  At the dip
bottom the observed 1--10 keV flux is $\sim$ 2/3 of the pre-dip flux,
and  the  covering  fraction  is  compatible  with  being  total.   No
substructures  are  present   within  the  dip,  which   points  to  a
homogeneous absorber.

If we consider the  orbital period of GX 13+1 ($\sim$  24.5 d) and the
possible  mass  range  for  the   companion  star  (between  1  and  5
M$_{\sun}$),  we can  estimate the  Roche lobe  radius of  the primary
according to the Eggleton formula \citep{eggleton83}:

\begin{equation}
\label{eggleton}
R_{RL} = a \frac{0.49 q^{2/3}}{0.6 q^{2/3} + ln(1 + q^{1/3})},
\end{equation}

where $q=M_{NS}/M_{comp}$ is the mass ratio, $a$ is orbital semi-major
axis, and  the mass  of the  NS is  assumed to  be 1.4  M$_{sun}$.  We
obtained  Roche-lobe  radii  in  the  (1.2--3.3)$\times$10$^{12}$  cm
range.  If the truncation radius ($R_{\textrm{tr}}$) for the accretion
disk is taken to be $\sim$  80\% of the Roche-lobe radius, we estimate
$R_{\textrm{tr}}$= (0.96--2.64)$\times$10$^{12}$  cm.  From  the same
calculations, we  derived that  the angle  subtended by  the companion
star's  radius is  $\sim$ 20\degr,  assuming it  completely fills  its
Roche lobe.  Because eclipses are absent, the inclination angle of the
system must be $\lesssim$ 70\degr.
 
If  matter  causing the  dip  is  assumed  for  simplicity to  have  a
spherical shape of  diameter $D_{blob}$ and fixed in  position, we can
estimate its diameter from the  duration of the dip (we conservatively
assumed as  dip duration  the entire dip  episode, shoulders  and deep
dip, of 1400 s) and the orbital period

\begin{equation}
D_{blob} = \frac{\Delta T_{dip}}{P_{orb}} \times R_{\textrm{tr}},
\end{equation}

which gives $D_{blob}$ = (6.3--17.4) $\times$ 10$^{8}$ cm. 

If the accretion disk height-to-radius ratio  were close to 0.1, as in
the standard  $\alpha$-disk case,  we immediately  note that  the disk
height is about three orders of magnitude greater than the blob radius
(as calculated  from a  variety of different  possible prescriptions),
and this low ratio makes it unlikely  that the blob is associated to a
thickening of  the disk  or some unstable  local structure  because in
this  case  we  would  expect  strong  variability  and  multiple  dip
substructures which are not  observed.  Alternatively, we might assume
that only a tiny  fraction of the top of the  bulge causes the shallow
dip, and thus  the peculiarity of this  source might be that  it is an
almost limiting case among the high-inclination systems, which is also
supported by the  general low upper value on the  inclination from the
eclipse absence.

We have shown that during the dip the observed hardening is mostly due
to a rising of a neutral  or very mildly ionized absorption column.  A
change  of \textit{only}  the warm  absorber characteristics,  with an
increase in the column density and a decrease of the ionization state,
does not  lead to  acceptable fit  results.  If  we consider  that the
column  density  due  to  the  ISM, as  derived  from  the  out-of-dip
spectrum, is $\sim$ 4.2$\times$10$^{22}$  cm$^{-2}$, at the dip bottom
we find an intervening local  column density $\Delta N_{\textrm H}$ of
4.5$\times$10$^{22}$  atoms cm$^{-2}$.   If  the  occulting region  is
placed at the minimum outer disk radius (10$^{12}$ cm) and the density
$n_{cold}$ is on  the order of $\Delta  N_{\textrm H}/D_{blob}$ $\sim$
3.7 $\times$ 10$^{13}$ cm$^{-3}$,  calculated using the blob diameter,
we  derive  a  maximum  ionization parameter  $\xi$  =  $L_x/(n_{cold}
r_{min}^2)$ $\sim$ 3.  Thus we conclude that the general picture of an
occulting, almost neutral  blob, set at a distance equal  to the outer
disk  radius is  self-consistent with  a  plasma in  a low  ionization
state.

\subsection*{GX 13+1 in context}

The nature of the dipping  phenomenon is still not firmly established,
neither is the physical connection with the geometry of the system. GX
13+1 is the most luminous accreting dipping system, and its broad-band
spectrum  closely  resembles the  typical  softness  of the  so-called
Z-sources and bright atoll group  \citep{dai09, disalvo09}, with a hot
($\sim$  1  keV) disk  spectrum  and  an optically  thick  Comptonized
component with a low ($\sim$ 3 keV) electron temperature.

In  addition to  the high  luminosity, GX  13+1 also  has the  longest
orbital period and hence the  largest outer-disk radius. We have shown
that if  the dipping  is caused  by a stable  structure at  the disk's
outer rim, even this intense flux is not able to strongly ionize it. X
1624-490  is the  second  dipping source  for  brightness and  longest
orbital period ($L_x$  $\sim$ 0.25 L$_{Edd}$, P$_{orb}$  21 hr), which
also  shows  a contribution  from  neutral  absorption during  dipping
\citep{diaztrigo06}. In this source an equal and large (a factor of 6)
increase both in the column density  of the ionized and neutral matter
has  been observed.  The  short  duration of  the  dip  and the  small
$Chandra$ effective area compared with  $XMM-Newton$ does not allow us
to constrain  the variability  of the  warm absorber  properties well,
which seems to  be modest compared with the significant  change of the
column  density  of  the  cold  absorber, however.  For  GX  13+1  the
ubiquitous presence  of the warm  absorber and the  possible formation
region  at  a  distance  of  $\sim$ 10$^{11}$  cm  suggest  that  cold
absorption (possibly located at a  distance $>$ 10$^{12}$ cm) and warm
absorption (disk-wind) are not  physically connected. New observations
with  higher statistics  will eventually  provide more  constraints on
this matter.

\section{Conclusions}

We have  reported the first spectroscopic  time-resolved investigation
of the periodic  dip of GX 13+1. The broad-band  spectrum derived by a
combined fit of \textit{Chandra}/HEG  and \textit{RXTE}/PCA allowed us
to consistently determine the continuum and discrete emission features
of the source.  \textit{Chandra} data confirm an out-flowing optically
thick warm  absorber.  Because of  the short  duration of the  dip, we
were unable to firmly constrain  possible changes in the properties of
the absorbing wind.  The observed spectral hardening during the dip is
mostly due to an increase in the column density of a neutral absorber,
while the warm-absorber component is  not modified with respect to the
out-of-dip  spectrum.   Simple  estimates  on the  dimensions  of  the
structure that  cause the dip  indicate a very small  occulting region
when compared  with the  expected scale-heights  at the  outer radius,
while  simple  geometric  considerations  on the  system  point  to  a
possible inclination of $\lesssim$ 70\degr.

\begin{acknowledgements} 
The authors thank  the anonymous referee for the  helpful comments and
suggestions. 

A.\ D.\ thanks M.\ Hanke for useful discussions about the
$simple\_gpile(2)$ model  and the  use of ISIS.\\  Authors acknowledge
support from  INAF/PRIN 2012-06.\\  A.\ D.\ , T.\ D.\ , R.\ I.\ N.\ R.\ acknowledge
support  from   Fondo  Finalizzato  alla  Ricerca   2012/13  from  the
University  of Palermo.   A.\ R.\ gratefully  acknowledges the  Sardinia
Regional  Government  for  the  financial  support  (P.\ O.\ R.\ Sardegna
F.S.E.  Operational Programme  of the  Autonomous Region  of Sardinia,
European Social  Fund 2007-2013 -  Axis IV Human  Resources, Objective
l.3, Line of Activity l.3.1). Work in Cagliari was partially funded by
the  Regione   Autonoma  della   Sardegna  through   POR-FSE  Sardegna
2007-2013,  L.R. 7/2007,  Progetti  di ricerca  di  base e  orientata,
Project N CRP-60529.\\

This research has made use of a collection of ISIS scripts provided by
the    Dr.\   Karl    Remeis   observatory,    Bamberg,   Germany    at
\url{http://www.sternwarte.uni-erlangen.de/isis/}.

\end{acknowledgements}
\bibliographystyle{aa}
\bibliography{refs}

\begin{thebibliography}{69}
\expandafter\ifx\csname natexlab\endcsname\relax\def\natexlab#1{#1}\fi

\bibitem[{{Asplund} {et~al.}(2009){Asplund}, {Grevesse}, {Sauval}, \&
  {Scott}}]{asplund09}
{Asplund}, M., {Grevesse}, N., {Sauval}, A.~J., \& {Scott}, P. 2009, \araa, 47,
  481

\bibitem[{{Ba{\l}uci{\'n}ska-Church} {et~al.}(1999){Ba{\l}uci{\'n}ska-Church},
  {Church}, {Oosterbroek}, {Segreto}, {Morley}, \& {Parmar}}]{balucinska99}
{Ba{\l}uci{\'n}ska-Church}, M., {Church}, M.~J., {Oosterbroek}, T., {et~al.}
  1999, \aap, 349, 495

\bibitem[{{Ba{\l}uci{\'n}ska-Church} {et~al.}(2004){Ba{\l}uci{\'n}ska-Church},
  {Church}, \& {Smale}}]{balucinska04}
{Ba{\l}uci{\'n}ska-Church}, M., {Church}, M.~J., \& {Smale}, A.~P. 2004,
  \mnras, 347, 334

\bibitem[{{Ba{\l}uci{\'n}ska-Church} {et~al.}(2011){Ba{\l}uci{\'n}ska-Church},
  {Schulz}, {Wilms}, {Gibiec}, {Hanke}, {Spencer}, {Rushton}, \&
  {Church}}]{balucinska11}
{Ba{\l}uci{\'n}ska-Church}, M., {Schulz}, N.~S., {Wilms}, J., {et~al.} 2011,
  \aap, 530, A102

\bibitem[{{Bandyopadhyay} {et~al.}(1999){Bandyopadhyay}, {Shahbaz}, {Charles},
  \& {Naylor}}]{bandyopadhyay99}
{Bandyopadhyay}, R.~M., {Shahbaz}, T., {Charles}, P.~A., \& {Naylor}, T. 1999,
  \mnras, 306, 417

\bibitem[{{Bhattacharyya} {et~al.}(2006){Bhattacharyya}, {Strohmayer},
  {Markwardt}, \& {Swank}}]{bhattacharyya06}
{Bhattacharyya}, S., {Strohmayer}, T.~E., {Markwardt}, C.~B., \& {Swank}, J.~H.
  2006, \apjl, 639, L31

\bibitem[{{Boirin} {et~al.}(2005){Boirin}, {M{\'e}ndez}, {D{\'{\i}}az Trigo},
  {Parmar}, \& {Kaastra}}]{boirin05}
{Boirin}, L., {M{\'e}ndez}, M., {D{\'{\i}}az Trigo}, M., {Parmar}, A.~N., \&
  {Kaastra}, J.~S. 2005, \aap, 436, 195

\bibitem[{{Boirin} {et~al.}(2004){Boirin}, {Parmar}, {Barret}, {Paltani}, \&
  {Grindlay}}]{boirin04}
{Boirin}, L., {Parmar}, A.~N., {Barret}, D., {Paltani}, S., \& {Grindlay},
  J.~E. 2004, \aap, 418, 1061

\bibitem[{{Church} \& {Ba{\l}uci{\'n}ska-Church}(2004)}]{church04}
{Church}, M.~J. \& {Ba{\l}uci{\'n}ska-Church}, M. 2004, \mnras, 348, 955

\bibitem[{{Church} {et~al.}(2012){Church}, {Gibiec},
  {Ba{\l}uci{\'n}ska-Church}, \& {Jackson}}]{church12}
{Church}, M.~J., {Gibiec}, A., {Ba{\l}uci{\'n}ska-Church}, M., \& {Jackson},
  N.~K. 2012, \aap, 546, A35

\bibitem[{{Corbet} {et~al.}(2010){Corbet}, {Pearlman}, {Buxton}, \&
  {Levine}}]{corbet10}
{Corbet}, R.~H.~D., {Pearlman}, A.~B., {Buxton}, M., \& {Levine}, A.~M. 2010,
  \apj, 719, 979

\bibitem[{{D'A{\`i}} {et~al.}(2010){D'A{\`i}}, {di Salvo}, {Ballantyne},
  {Iaria}, {Robba}, {Papitto}, {Riggio}, {Burderi}, {Piraino}, {Santangelo},
  {Matt}, {Dov{\v c}iak}, \& {Karas}}]{dai10}
{D'A{\`i}}, A., {di Salvo}, T., {Ballantyne}, D., {et~al.} 2010, \aap, 516, A36

\bibitem[{{D'A{\'{\i}}} {et~al.}(2007){D'A{\'{\i}}}, {Iaria}, {Di Salvo},
  {Lavagetto}, \& {Robba}}]{dai07a}
{D'A{\'{\i}}}, A., {Iaria}, R., {Di Salvo}, T., {Lavagetto}, G., \& {Robba},
  N.~R. 2007, \apj, 671, 2006

\bibitem[{{D'A{\`i}} {et~al.}(2009){D'A{\`i}}, {Iaria}, {Di Salvo}, {Matt}, \&
  {Robba}}]{dai09}
{D'A{\`i}}, A., {Iaria}, R., {Di Salvo}, T., {Matt}, G., \& {Robba}, N.~R.
  2009, \apjl, 693, L1

\bibitem[{{Di Salvo} {et~al.}(2009){Di Salvo}, {D'A{\'{\i}}}, {Iaria},
  {Burderi}, {Dov{\v c}iak}, {Karas}, {Matt}, {Papitto}, {Piraino}, {Riggio},
  {Robba}, \& {Santangelo}}]{disalvo09}
{Di Salvo}, T., {D'A{\'{\i}}}, A., {Iaria}, R., {et~al.} 2009, \mnras, 398,
  2022

\bibitem[{{Di Salvo} {et~al.}(2002){Di Salvo}, {Farinelli}, {Burderi},
  {Frontera}, {Kuulkers}, {Masetti}, {Robba}, {Stella}, \& {van der
  Klis}}]{disalvo02}
{Di Salvo}, T., {Farinelli}, R., {Burderi}, L., {et~al.} 2002, \aap, 386, 535

\bibitem[{{D{\'{\i}}az Trigo} {et~al.}(2006){D{\'{\i}}az Trigo}, {Parmar},
  {Boirin}, {M{\'e}ndez}, \& {Kaastra}}]{diaztrigo06}
{D{\'{\i}}az Trigo}, M., {Parmar}, A.~N., {Boirin}, L., {M{\'e}ndez}, M., \&
  {Kaastra}, J.~S. 2006, \aap, 445, 179

\bibitem[{{D{\'{\i}}az Trigo} {et~al.}(2009){D{\'{\i}}az Trigo}, {Parmar},
  {Boirin}, {Motch}, {Talavera}, \& {Balman}}]{diaztrigo09}
{D{\'{\i}}az Trigo}, M., {Parmar}, A.~N., {Boirin}, L., {et~al.} 2009, \aap,
  493, 145

\bibitem[{{D{\'{\i}}az Trigo} {et~al.}(2012){D{\'{\i}}az Trigo}, {Sidoli},
  {Boirin}, \& {Parmar}}]{diaztrigo12}
{D{\'{\i}}az Trigo}, M., {Sidoli}, L., {Boirin}, L., \& {Parmar}, A.~N. 2012,
  \aap, 543, A50

\bibitem[{{Eggleton}(1983)}]{eggleton83}
{Eggleton}, P.~P. 1983, \apj, 268, 368

\bibitem[{{Frank} {et~al.}(1987){Frank}, {King}, \& {Lasota}}]{frank87}
{Frank}, J., {King}, A.~R., \& {Lasota}, J.-P. 1987, \aap, 178, 137

\bibitem[{{Gavriil} {et~al.}(2012){Gavriil}, {Strohmayer}, \&
  {Bhattacharyya}}]{gavriil12}
{Gavriil}, F.~P., {Strohmayer}, T.~E., \& {Bhattacharyya}, S. 2012, \apj, 753,
  2

\bibitem[{{Gris{\'e}} {et~al.}(2013){Gris{\'e}}, {Kaaret}, {Corbel}, {Cseh}, \&
  {Feng}}]{grise13}
{Gris{\'e}}, F., {Kaaret}, P., {Corbel}, S., {Cseh}, D., \& {Feng}, H. 2013,
  \mnras, 433, 1023

\bibitem[{{Hanke} {et~al.}(2009){Hanke}, {Wilms}, {Nowak}, {Pottschmidt},
  {Schulz}, \& {Lee}}]{hanke09}
{Hanke}, M., {Wilms}, J., {Nowak}, M.~A., {et~al.} 2009, \apj, 690, 330

\bibitem[{{Homan} {et~al.}(2005){Homan}, {Miller}, {Wijnands}, {van der Klis},
  {Belloni}, {Steeghs}, \& {Lewin}}]{homan05}
{Homan}, J., {Miller}, J.~M., {Wijnands}, R., {et~al.} 2005, \apj, 623, 383

\bibitem[{{Homan} {et~al.}(2004){Homan}, {Wijnands}, {Rupen}, {Fender},
  {Hjellming}, {di Salvo}, \& {van der Klis}}]{homan04}
{Homan}, J., {Wijnands}, R., {Rupen}, M.~P., {et~al.} 2004, \aap, 418, 255

\bibitem[{{Houck}(2002)}]{houck02}
{Houck}, J.~C. 2002, in High Resolution X-ray Spectroscopy with XMM-Newton and
  Chandra, ed. G.~{Branduardi-Raymont}

\bibitem[{{Hyodo} {et~al.}(2009){Hyodo}, {Ueda}, {Yuasa}, {Maeda}, {Makishima},
  \& {Koyama}}]{hyodo09}
{Hyodo}, Y., {Ueda}, Y., {Yuasa}, T., {et~al.} 2009, \pasj, 61, 99

\bibitem[{{Iaria} {et~al.}(2014){Iaria}, {Di Salvo}, {Burderi}, {Riggio},
  {D'A{\`i}}, \& {Robba}}]{iaria14}
{Iaria}, R., {Di Salvo}, T., {Burderi}, L., {et~al.} 2014, \aap, 561, A99

\bibitem[{{Iaria} {et~al.}(2013){Iaria}, {Di Salvo}, {D'A{\`i}}, {Burderi},
  {Mineo}, {Riggio}, {Papitto}, \& {Robba}}]{iaria13}
{Iaria}, R., {Di Salvo}, T., {D'A{\`i}}, A., {et~al.} 2013, \aap, 549, A33

\bibitem[{{Iaria} {et~al.}(2007){Iaria}, {Lavagetto}, {D'A{\'{\i}}}, {di
  Salvo}, \& {Robba}}]{iaria07}
{Iaria}, R., {Lavagetto}, G., {D'A{\'{\i}}}, A., {di Salvo}, T., \& {Robba},
  N.~R. 2007, \aap, 463, 289

\bibitem[{{in't Zand} {et~al.}(2003){in't Zand}, {Hulleman}, {Markwardt},
  {M{\'e}ndez}, {Kuulkers}, {Cornelisse}, {Heise}, {Strohmayer}, \&
  {Verbunt}}]{intzand03}
{in't Zand}, J.~J.~M., {Hulleman}, F., {Markwardt}, C.~B., {et~al.} 2003, \aap,
  406, 233

\bibitem[{{Kallman} {et~al.}(2004){Kallman}, {Palmeri}, {Bautista}, {Mendoza},
  \& {Krolik}}]{kallman04}
{Kallman}, T.~R., {Palmeri}, P., {Bautista}, M.~A., {Mendoza}, C., \& {Krolik},
  J.~H. 2004, \apjs, 155, 675

\bibitem[{{Krolik} {et~al.}(1981){Krolik}, {McKee}, \& {Tarter}}]{krolik81}
{Krolik}, J.~H., {McKee}, C.~F., \& {Tarter}, C.~B. 1981, \apj, 249, 422

\bibitem[{{Kuulkers} {et~al.}(2013){Kuulkers}, {Kouveliotou}, {Belloni},
  {Cadolle Bel}, {Chenevez}, {D{\'{\i}}az Trigo}, {Homan}, {Ibarra}, {Kennea},
  {Mu{\~n}oz-Darias}, {Ness}, {Parmar}, {Pollock}, {van den Heuvel}, \& {van
  der Horst}}]{kuulkers13}
{Kuulkers}, E., {Kouveliotou}, C., {Belloni}, T., {et~al.} 2013, \aap, 552, A32

\bibitem[{{Kuulkers} {et~al.}(1998){Kuulkers}, {Wijnands}, {Belloni}, {Mendez},
  {van der Klis}, \& {van Paradijs}}]{kuulkers98}
{Kuulkers}, E., {Wijnands}, R., {Belloni}, T., {et~al.} 1998, \apj, 494, 753

\bibitem[{{Lee} {et~al.}(2002){Lee}, {Reynolds}, {Remillard}, {Schulz},
  {Blackman}, \& {Fabian}}]{lee02}
{Lee}, J.~C., {Reynolds}, C.~S., {Remillard}, R., {et~al.} 2002, \apj, 567,
  1102

\bibitem[{{Lubow}(1989)}]{lubow89}
{Lubow}, S.~H. 1989, \apj, 340, 1064

\bibitem[{{Mainardi} {et~al.}(2010){Mainardi}, {Paizis}, {Farinelli},
  {Kuulkers}, {Rodriguez}, {Hannikainen}, {Savolainen}, {Piraino}, {Bazzano},
  \& {Santangelo}}]{mainardi10}
{Mainardi}, L.~I., {Paizis}, A., {Farinelli}, R., {et~al.} 2010, \aap, 512, A57

\bibitem[{{Mason}(1986)}]{mason86}
{Mason}, K.~O. 1986, in Lecture Notes in Physics, Berlin Springer Verlag, Vol.
  266, The Physics of Accretion onto Compact Objects, ed. K.~O. {Mason}, M.~G.
  {Watson}, \& N.~E. {White}, 29

\bibitem[{{Matsuba} {et~al.}(1995){Matsuba}, {Dotani}, {Mitsuda}, {Asai},
  {Lewin}, {van Paradijs}, \& {van der Klis}}]{matsuba95}
{Matsuba}, E., {Dotani}, T., {Mitsuda}, K., {et~al.} 1995, \pasj, 47, 575

\bibitem[{{Miller} {et~al.}(2006){Miller}, {Raymond}, {Homan}, {Fabian},
  {Steeghs}, {Wijnands}, {Rupen}, {Charles}, {van der Klis}, \&
  {Lewin}}]{miller06}
{Miller}, J.~M., {Raymond}, J., {Homan}, J., {et~al.} 2006, \apj, 646, 394

\bibitem[{{Naik} {et~al.}(2001){Naik}, {Agrawal}, {Rao}, {Paul}, {Seetha}, \&
  {Kasturirangan}}]{naik01}
{Naik}, S., {Agrawal}, P.~C., {Rao}, A.~R., {et~al.} 2001, \apj, 546, 1075

\bibitem[{{Nowak} {et~al.}(2008){Nowak}, {Juett}, {Homan}, {Yao}, {Wilms},
  {Schulz}, \& {Canizares}}]{nowak08}
{Nowak}, M.~A., {Juett}, A., {Homan}, J., {et~al.} 2008, \apj, 689, 1199

\bibitem[{{Oosterbroek} {et~al.}(2001){Oosterbroek}, {Parmar}, {Sidoli}, {in't
  Zand}, \& {Heise}}]{oosterbroek01}
{Oosterbroek}, T., {Parmar}, A.~N., {Sidoli}, L., {in't Zand}, J.~J.~M., \&
  {Heise}, J. 2001, \aap, 376, 532

\bibitem[{{Paizis} {et~al.}(2006){Paizis}, {Farinelli}, {Titarchuk},
  {Courvoisier}, {Bazzano}, {Beckmann}, {Frontera}, {Goldoni}, {Kuulkers},
  {Mereghetti}, {Rodriguez}, \& {Vilhu}}]{paizis06}
{Paizis}, A., {Farinelli}, R., {Titarchuk}, L., {et~al.} 2006, \aap, 459, 187

\bibitem[{{Parmar} {et~al.}(1989){Parmar}, {Gottwald}, {van der Klis}, \& {van
  Paradijs}}]{parmar89}
{Parmar}, A.~N., {Gottwald}, M., {van der Klis}, M., \& {van Paradijs}, J.
  1989, \apj, 338, 1024

\bibitem[{{Parmar} {et~al.}(1986){Parmar}, {White}, {Giommi}, \&
  {Gottwald}}]{parmar86}
{Parmar}, A.~N., {White}, N.~E., {Giommi}, P., \& {Gottwald}, M. 1986, \apj,
  308, 199

\bibitem[{{Ponti} {et~al.}(2012){Ponti}, {Fender}, {Begelman}, {Dunn},
  {Neilsen}, \& {Coriat}}]{ponti12}
{Ponti}, G., {Fender}, R.~P., {Begelman}, M.~C., {et~al.} 2012, \mnras, 422,
  L11

\bibitem[{{Popham} \& {Sunyaev}(2001)}]{popham01}
{Popham}, R. \& {Sunyaev}, R. 2001, \apj, 547, 355

\bibitem[{{Pozdnyakov} {et~al.}(1983){Pozdnyakov}, {Sobol}, \&
  {Syunyaev}}]{pozdnyakov83}
{Pozdnyakov}, L.~A., {Sobol}, I.~M., \& {Syunyaev}, R.~A. 1983, Astrophysics
  and Space Physics Reviews, 2, 189

\bibitem[{{Shidatsu} {et~al.}(2013){Shidatsu}, {Ueda}, {Nakahira}, {Done},
  {Morihana}, {Sugizaki}, {Mihara}, {Hori}, {Negoro}, {Kawai}, {Yamaoka},
  {Ebisawa}, {Matsuoka}, {Serino}, {Yoshikawa}, {Nagayama}, \&
  {Matsunaga}}]{shidatsu13}
{Shidatsu}, M., {Ueda}, Y., {Nakahira}, S., {et~al.} 2013, ArXiv e-prints

\bibitem[{{Shirey} {et~al.}(1999){Shirey}, {Levine}, \& {Bradt}}]{shirey99}
{Shirey}, R.~E., {Levine}, A.~M., \& {Bradt}, H.~V. 1999, \apj, 524, 1048

\bibitem[{{Sidoli} {et~al.}(2001){Sidoli}, {Oosterbroek}, {Parmar}, {Lumb}, \&
  {Erd}}]{sidoli01}
{Sidoli}, L., {Oosterbroek}, T., {Parmar}, A.~N., {Lumb}, D., \& {Erd}, C.
  2001, \aap, 379, 540

\bibitem[{{Sidoli} {et~al.}(2002){Sidoli}, {Parmar}, {Oosterbroek}, \&
  {Lumb}}]{sidoli02}
{Sidoli}, L., {Parmar}, A.~N., {Oosterbroek}, T., \& {Lumb}, D. 2002, \aap,
  385, 940

\bibitem[{{Smale} {et~al.}(2001){Smale}, {Church}, \&
  {Ba{\l}uci{\'n}ska-Church}}]{smale01}
{Smale}, A.~P., {Church}, M.~J., \& {Ba{\l}uci{\'n}ska-Church}, M. 2001, \apj,
  550, 962

\bibitem[{{Smale} {et~al.}(2002){Smale}, {Church}, \&
  {Ba{\l}uci{\'n}ska-Church}}]{smale02}
{Smale}, A.~P., {Church}, M.~J., \& {Ba{\l}uci{\'n}ska-Church}, M. 2002, \apj,
  581, 1286

\bibitem[{{Smale} \& {Wachter}(1999)}]{smale99}
{Smale}, A.~P. \& {Wachter}, S. 1999, \apj, 527, 341

\bibitem[{{Titarchuk}(1994)}]{titarchuk94}
{Titarchuk}, L. 1994, \apj, 434, 570

\bibitem[{{Ueda} {et~al.}(2001){Ueda}, {Asai}, {Yamaoka}, {Dotani}, \&
  {Inoue}}]{ueda01}
{Ueda}, Y., {Asai}, K., {Yamaoka}, K., {Dotani}, T., \& {Inoue}, H. 2001,
  \apjl, 556, L87

\bibitem[{{Ueda} {et~al.}(1998){Ueda}, {Inoue}, {Tanaka}, {Ebisawa}, {Nagase},
  {Kotani}, \& {Gehrels}}]{ueda98}
{Ueda}, Y., {Inoue}, H., {Tanaka}, Y., {et~al.} 1998, \apj, 492, 782

\bibitem[{{Ueda} {et~al.}(2005){Ueda}, {Mitsuda}, {Murakami}, \&
  {Matsushita}}]{ueda05}
{Ueda}, Y., {Mitsuda}, K., {Murakami}, H., \& {Matsushita}, K. 2005, \apj, 620,
  274

\bibitem[{{Ueda} {et~al.}(2004){Ueda}, {Murakami}, {Yamaoka}, {Dotani}, \&
  {Ebisawa}}]{ueda04}
{Ueda}, Y., {Murakami}, H., {Yamaoka}, K., {Dotani}, T., \& {Ebisawa}, K. 2004,
  \apj, 609, 325

\bibitem[{{Uzdensky}(2013)}]{uzdensky13}
{Uzdensky}, D.~A. 2013, \apj, 775, 103

\bibitem[{{van Peet} {et~al.}(2009){van Peet}, {Costantini}, {M{\'e}ndez},
  {Paerels}, \& {Cottam}}]{vanpeet09}
{van Peet}, J.~C.~A., {Costantini}, E., {M{\'e}ndez}, M., {Paerels}, F.~B.~S.,
  \& {Cottam}, J. 2009, \aap, 497, 805

\bibitem[{{Verner} {et~al.}(1996){Verner}, {Ferland}, {Korista}, \&
  {Yakovlev}}]{verner96}
{Verner}, D.~A., {Ferland}, G.~J., {Korista}, K.~T., \& {Yakovlev}, D.~G. 1996,
  \apj, 465, 487

\bibitem[{{White} \& {Holt}(1982)}]{white82}
{White}, N.~E. \& {Holt}, S.~S. 1982, \apj, 257, 318

\bibitem[{{Younes} {et~al.}(2009){Younes}, {Boirin}, \& {Sabra}}]{younes09}
{Younes}, G., {Boirin}, L., \& {Sabra}, B. 2009, \aap, 502, 905

\bibitem[{{Zimmerman} {et~al.}(2005){Zimmerman}, {Narayan}, {McClintock}, \&
  {Miller}}]{zimmerman05}
{Zimmerman}, E.~R., {Narayan}, R., {McClintock}, J.~E., \& {Miller}, J.~M.
  2005, \apj, 618, 832

\end{thebibliography}
%
\end{document}